\documentclass[sigconf]{acmart}
\usepackage{booktabs}
\usepackage{multirow}
\usepackage[normalem]{ulem}
\useunder{\uline}{\ul}{}
\usepackage{subcaption}
\usepackage{graphicx}

\usepackage{amsmath,amsfonts,bm}

\def\eqref#1{equation~\ref{#1}}

\def\1{\bm{1}}

\def\vb{{\bm{b}}}

\def\ve{{\bm{e}}}
\def\vf{{\bm{f}}}

\def\vh{{\bm{h}}}

\def\vp{{\bm{p}}}

\def\vr{{\bm{r}}}

\def\mE{{\bm{E}}}

\def\mH{{\bm{H}}}

\def\mK{{\bm{K}}}

\def\mP{{\bm{P}}}
\def\mQ{{\bm{Q}}}
\def\mR{{\bm{R}}}

\def\mV{{\bm{V}}}
\def\mW{{\bm{W}}}

\DeclareMathAlphabet{\mathsfit}{\encodingdefault}{\sfdefault}{m}{sl}
\SetMathAlphabet{\mathsfit}{bold}{\encodingdefault}{\sfdefault}{bx}{n}

\def\gS{{\mathcal{S}}}

\def\gU{{\mathcal{U}}}
\def\gV{{\mathcal{V}}}

\def\sR{{\mathbb{R}}}
\def\sS{{\mathbb{S}}}

\AtBeginDocument{%
  \providecommand\BibTeX{{%
    \normalfont B\kern-0.5em{\scshape i\kern-0.25em b}\kern-0.8em\TeX}}}

\copyrightyear{2021} 
\acmYear{2021} 
\setcopyright{acmcopyright}\acmConference[CIKM '21]{Proceedings of the 30th ACM International Conference on Information and Knowledge Management}{November 1--5, 2021}{Virtual Event, QLD, Australia}
\acmBooktitle{Proceedings of the 30th ACM International Conference on Information and Knowledge Management (CIKM '21), November 1--5, 2021, Virtual Event, QLD, Australia}
\acmPrice{15.00}
\acmDOI{10.1145/3459637.3482448}
\acmISBN{978-1-4503-8446-9/21/11}

\begin{document}
\fancyhead{}

\title{Lightweight Self-Attentive Sequential Recommendation}

\author{Yang Li}
\affiliation{%
  \institution{The University of Queensland}
}
\email{yang.li@uq.edu.au}

\author{Tong Chen}
\affiliation{%
  \institution{The University of Queensland}
}
\email{tong.chen@uq.edu.au	}

\author{Peng-Fei Zhang}
\affiliation{%
  \institution{The University of Queensland}
}
\email{mima.zpf@gmail.com}

\author{Hongzhi Yin}
\affiliation{%
  \institution{The University of Queensland}
}
\email{h.yin1@uq.edu.au	}

\begin{abstract}
  Modern deep neural networks (DNNs) have greatly facilitated the development of sequential recommender systems by achieving state-of-the-art recommendation performance on various sequential recommendation tasks. Given a sequence of interacted items, existing DNN-based sequential recommenders commonly embed each item into a unique vector to support subsequent computations of the user interest. However, due to the potentially large number of items, the over-parameterised item embedding matrix of a sequential recommender has become a memory bottleneck for efficient deployment in resource-constrained environments, e.g., smartphones and other edge devices. Furthermore, we observe that the widely-used multi-head self-attention, though being effective in modelling sequential dependencies among items, heavily relies on redundant attention units to fully capture both global and local item-item transition patterns within a sequence.
  
  In this paper, we introduce a novel lightweight self-attentive network (LSAN) for sequential recommendation. To aggressively compress the original embedding matrix, LSAN leverages the notion of compositional embeddings, where each item embedding is composed by merging a group of selected base embedding vectors derived from substantially smaller embedding matrices. Meanwhile, to account for the intrinsic dynamics of each item, we further propose a temporal context-aware embedding composition scheme. Besides, we develop an innovative twin-attention network that alleviates the redundancy of the traditional multi-head self-attention while retaining full capacity for capturing long- and short-term (i.e., global and local) item dependencies. Comprehensive experiments demonstrate that LSAN significantly advances the accuracy and memory efficiency of existing sequential recommenders.
\end{abstract}

\maketitle

\section{Introduction}\label{sec:intro}
Modelling sequential user behaviours has received great attention in contemporary applications, such as e-commerce, online services, and smart transport \cite{YinC16}. Among these applications, sequential recommender systems (SRSs) have become a prominent solution to information overload on the web. The main goal of SRSs is to make a proactive recommendation on the next item a user may be interested in by mining the user's recent preferences from the sequence of her/his interacted items.

Early SRSs incorporate Markov chain-based models \cite{RendleFS10,ChengYLK13,HeKM17} to capture high-order sequential patterns based on Markov chain (MC), which essentially factorises a user-specific item-item transition tensor by considering first-order Markov chain. However, these methods primarily learn the transition patterns based on the most recent item interactions, neglecting long-term (i.e., global) user preferences. With the revolution of deep neural networks (DNNs), various deep methods have been proposed for the sequential recommendation \cite{QiuLHY19,QiuYHC20,QiuHLY20,QiuPosRec21,LiCLYH21}, especially recurrent neural network-based \cite{chen2018sequential,LiLZSC19,chen2019air,zheng2020sentiment,ChenYNP0020} sequential recommenders. Notably, the majority of state-of-the-art SRSs are latent factor models, where each item is mapped into a unique vector representation (a.k.a. embedding), and the item embedding is then used to calculate the sequential preferences of the target user.

With the fast pace of digitisation and hardware revolution, there has been a recent surge of moving data analytics from cloud servers to edge devices \cite{shi2016edge} to ensure timeliness and privacy. As sequential recommendation involves frequent updates on a user's behaviour records, developing lightweight SRSs appears to be an ongoing trend, because such on-device computation capability can prevent potential latency caused by communications with the cloud and effectively retains users' personal data on their own devices. Due to the sheer volume of different items (e.g., Alibaba's billion-scale item set \cite{wang2018billion}), the item embeddings in latent SRSs are the main source of memory consumption \cite{WangYCHWZH20} rather than other parameters like weights and biases of DNNs. In this regard, recent studies on lightweight recommenders \cite{ShiMNY20,LianWLLC020,liu2020automated,ChenYZHWW21} are predominantly focused on compressing the originally large item embedding matrix to improve the memory efficiency of recommenders. Their core idea of such compression is compositional embeddings, where a recommender consists of a small number (substantially smaller than the number of all items) of base embeddings, such that an item can be represented as a distinct combination of selected base embeddings. However, most compositional embedding-based recommenders are designed for static recommendation settings, where the recommendation results for each user are purely conditioned on the static user-item affinity instead of her/his interest dynamics. 

As a common practice, the aforementioned lightweight recommenders usually combine compositional embeddings with off-the-shelf deep recommendation modules (e.g., the DLRS in \cite{ShiMNY20} and the DeepFM in \cite{LiuGCJL21}). Though a similar solution can be sought for lightweight SRSs by straightforwardly feeding compositional item embeddings into sequential DNNs (e.g., recurrent neural networks), an over-parameterised network structure will in fact defeat the purpose of a memory-efficient model. Also, the excessive computations may impede the timeliness of a model's on-device inference. Hence, in addition to a lightly parameterised item embedding scheme, a lightweight SRS should also be able to thoroughly discover the temporal signals from all interacted items with a carefully designed, compact, yet effective sequence mining paradigm.

To this end, we propose \underline{l}ightweight \underline{s}elf-\underline{a}ttentive \underline{n}etwork (LSAN), a novel solution to memory-efficient sequential recommendation that simultaneously addresses those two key challenges. Specifically, LSAN aggressively replaces the item embedding matrix with $N$ base embedding matrices, each of which contains substantially fewer embedding vectors (i.e., base embeddings) than the total amount of items, i.e., $N \ll |\mathcal{V}|$ for item set $\mathcal{V}$. Then, compositional item embeddings are generated by fusing $N$ base embeddings respectively selected from each base embedding matrix. To ensure the uniqueness of each composed item embedding, we design a context-aware temporal compositional embedding scheme, where base embeddings are located via a quotient-remainder \cite{ShiMNY20} operation. Unlike traditional compositional item embeddings that stay fixed regardless of any temporal information in a sequence, we propose to dynamically alter the generated item embedding according to sequence-specific contexts by attentively merging the base embeddings for each item. The rationale is that in a sequence, every item's relevance is sensitive to factors like seasonal changes and adjacent items \cite{WuABSJ17,KumarZL19}, hence we let LSAN account for such information when generating compositional item embeddings. 

In LSAN, we resort to self-attention for modelling the temporal patterns among the interacted items. However, though the self-attention  \cite{LvJYSLYN19,KangM18,zhou2020s3} is widely acknowledged as a lightly-parameterised and effective approach for capturing sequential information in SRSs compared with the popular alternative -- RNNs, our observation is that recommenders using self-attention can still incur parameter redundancy. For sequential recommendation tasks, the attention module should be able to capture both long- and short-term (i.e., global and local) preferences of a user. Unfortunately, recent studies \cite{WuLLLH20,wu2019pay} point out that self-attention tends to over-emphasise local relationships between adjacent items, making it difficult to learn the correlations between items that are far from each other in a sequence. Hence, existing recommenders commonly employ multi-head self-attention, so as to enhance the modelling capacity and acquire sufficient global information. Such redundancy can be tolerated and may benefit the recommendation accuracy when the computing resource allows, but it fails to meet the highly constrained deployment environment in the context of memory-efficient SRSs. In light of this, we introduce a novel twin-attention paradigm in LSAN, where the global and local preference signals are separately captured via two specialised modules instead of a group of general attention units. As such, coupled with the dynamic compositional item embedding scheme, LSAN effectively learns local and global user preference signals for accurate sequential recommendation without the need for an excessively complex and large model. 

With the proposed LSAN, our main contributions to lightweight sequential recommendation are three-fold:
\begin{itemize}
    \item We devise a dynamic context-aware compositional embedding scheme, which largely decreases the memory footprint of item embedding matrix -- the major consumer of memory space of SRSs -- and ensures the uniqueness and dynamics of generated item embeddings at the same time.
    \item We propose a novel twin-attention sequential framework, which specialises the learning of long- and short-term user preference signals via a dedicated self-attention and convolution operation, respectively. This facilitates explicit modelling of both global and local patterns while avoiding the redundancy of multi-head self-attention modules.
    \item Extensive experiments are conducted on three  benchmark datasets. The results demonstrate the advantageous effectiveness and memory efficiency of LSAN against state-of-the-art baseline methods.
\end{itemize}

\section{Problem Formulation}
Let $\mathcal{V}$, $\mathcal{U}$ 
be the sets of items and users, respectively. 
We use $\gS_u=\{v_1, v_2, ..., v_{T}\}$ to denote a sequence of $T$ chronologically ordered items that user $u \in \gU$ has interacted with. Each item $v_i \in \gS_u$ is assigned an order index $i=1,2,...,T$ which reflects the position of an item in the sequence. To locate $v_i$ among the set of $|\mathcal{V}|$ items, we define a function $\text{index}(v_i)$ that maps $v_i$ to a unique and fixed global index $1,2,...,|\mathcal{V}|$. 
Then, given a sequence of interactions $\sS_u$, our goal is to compute a ranking list consisting of top $K$ items that $u$ is most likely to visit at the next time step $T + 1$.

\begin{figure*}
    \centering
    \includegraphics[width=\textwidth]{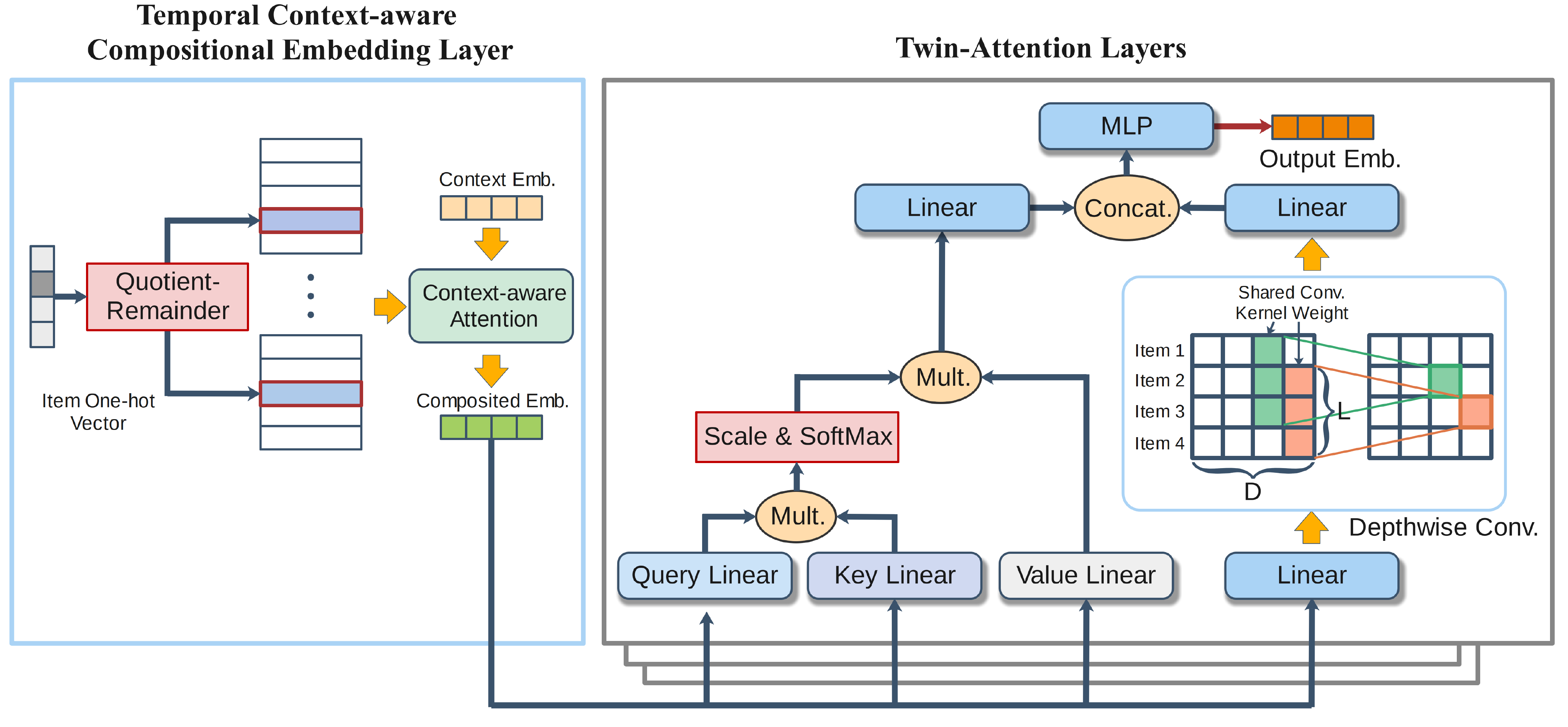}
    \caption{An overview of the proposed LSAN model.}
    \label{fig:overview}
\end{figure*}

\section{Methodology}
In this section, we introduce our proposed memory-efficient sequential recommender, namely LSAN. LSAN consists of two main components: (1) a dynamic context-aware compositional embedding layer that enables a lightweight yet highly expressive item embedding paradigm; and (2) a twin-attention network that effectively learns global and local user preferences without the need for redundant multi-head self-attention modules. In what follows, we present the design of LSAN in detail.

\subsection{Dynamic Context-aware Compositional Embedding}
\label{TACEN}
Recall that in a typical latent factor-based recommender system, each item $v_i$ is associated with a unique $D$-dimensional embedding vector, which corresponds to the $\index(v_i)$-th row in an embedding table $\mE \in \sR^{|\gV| \times D}$. The memory complexity of maintaining such an embedding table is $\mathcal{O}(|\gV|D)$, where the memory cost will become impractical for edge devices when $\gV$ is large-scale. To reduce the size of $\mE$ for better memory efficiency, we replace $\mE$ with a set of $N$ base embedding tables denoted as $\{\widetilde{\mE}_1,\widetilde{\mE}_2,...,\widetilde{\mE}_N\}$, where $\widetilde{\mE}_n \in \sR^{m_n \times D}$ for $n=1,2,...,N$. Here, $m_n$ indicates the number of base embeddings in the $n$-th base embedding table $\widetilde{\mE}_n$ and $m_n \ll |\gV|$. For each item, its compositional embedding is produced by first selecting one base embedding vector from each $\widetilde{\mE}_n$, and then attentively fusing all selected base embeddings into a single vector. It can be concluded that there are $\Pi^{N}_{n=1} m_n$ different combinations of base embeddings. Thus, we only need to make $\Pi^{N}_{n=1} m_n \geqslant |\gV|$ to ensure the uniqueness of constructed embeddings for each item. To guarantee that each item receives a distinct combination of base embeddings, we resort to the quotient-remainder trick \cite{ShiMNY20} that does not introduce any additional learnable parameters. Particularly, taking the first base embedding table $\widetilde{\mE}_1$ as an example, the corresponding base embedding index $q_i$ (i.e., the row index in $\widetilde{\mE}_1$) for item $v_i \in \mathcal{S}_u$ can be computed by a remainder function over the base embedding table size $m_1$:
\begin{equation}
    q_i = \text{index}(v_i) \bmod m_1.
\end{equation}
Then, the first base embedding $\widetilde{\ve}^{\,1}_{i}$ for $v_i$ can be retrieved by a look-up operation on $\widetilde{\mE}_1$ w.r.t. row index $q_i$. Mathematically, let $\vf_i \in \sR^{|\gV|}$ be the one-hot encoding of $v_i$, then a hash matrix $\mR^1 \in \sR^{m_1 \times |\gV|}$ for $\widetilde{\mE}_1$ can be computed element-wise via:
\begin{equation}
    \mR^1_{i, j}=\left\{\begin{array}{ll}
    1 & \text{ if } j \bmod m_1 = \text{index}(v_i) \\
    0 & \text{ otherwise }
    \end{array}\right..
\end{equation}
Therefore, the look-up operation of $v_i$'s first base embedding $\widetilde{\ve}^{\,1}_{i}$ can be mathematically formulated as:
\begin{equation}
    \widetilde{\ve}^{\,1}_{i} = {\widetilde{\mE}_1}^{\,\top} \mR^1 \vf_i.
\end{equation}
Analogously, for $n=2,3,...,N$, $v_i$'s hash matrix $\mR^n$ for the $n$-th base embedding table can be generalised as:
\begin{equation}
    \mR^n_{i, j}=\left\{\begin{array}{ll}
    1 & \text{ if } j \bmod m_j = \text{index}(v_i) \setminus \Pi^{n-1}_{n=1}m_n \\
    0 & \text{ otherwise }
    \end{array}\right.,
\end{equation}
where the index in the $n$-th base embedding table for item $v_i$ is determined by the resulting quotient from the prior base embedding tables, i.e., $\text{index}(v_i) \setminus \Pi^{n-1}_{n=1}m_n$. Then, we obtain the base embedding $\widetilde{\ve}^n_{i}$ from $\mR^n$ via:
\begin{equation}
    \widetilde{\ve}^n_{i} = \widetilde{\mE}_n \mR^n \vf_i.
\end{equation}
Through the quotient-remainder trick, we now have acquired a set of base embeddings $\{\widetilde{\ve}^{\,1}_{i},\widetilde{\ve}^{\,2}_{i},...,\widetilde{\ve}^{N}_{i}\}$ for item $v_i$. Intuitively, a unified item embedding can be easily formed by an ensemble operation, such as element-wise addition/multiplication. However, such constructed embeddings are fixed, and are insufficient in capturing the intrinsic dynamics of an item's properties. As pointed out by \cite{WuABSJ17,KumarZL19}, learning context-aware temporal item embeddings is beneficial for mining a user's preferences from her/his interaction sequences. In order to bring such temporal contexts into the generated compositional item embeddings, we propose to attentively assign different weights to the selected base embeddings conditioned on the context around the target item. Specifically, for each item $v_i \in \mathcal{S}_u$ ($i = 1, 2, ...,T$), we have a context $r_{i}$ representing this item's situation within the sequence. The construction of $r_{i}$ can be highly flexible, where in our work, based on the side information shared by all of our experimental datasets (see Section \ref{sec:exp}), we define $r_{i} = (c_{i-1}, c_{i}, \text{time}(i))$ as a triplet of the categories of the previous and current items and the discrete time slot (i.e., every hour of a day). We denote $R = \{r_1, r_2, r_{|R|}\}$ be the set of unique context tuples. Each tuple $r \in R$ is assigned with a one-hot vector. Then, we can map the one-hot encoding vector $r_{i}$ into a dense context embedding $\vr_{i} \in \sR^D$. Note that a padding label for the category information is adopted when $t=1$. Under a given context $\vr_{i}$, we can calculate an attention weight for each base embedding:
\begin{equation}
    \alpha_{n} = \frac{\exp(\vr_{i}^{\top}\operatorname{SiLU}(\mW_a \widetilde{\ve}^{n}_{i}))}{\sum^{N}_{n=1}\exp(\vr_{i}^{\top}\operatorname{SiLU}(\mW_a \widetilde{\ve}^{n}_{i}))},
\end{equation}
where $\operatorname{SiLU}(x) = x \cdot \operatorname{sigmoid}(x)$ is an activation function that is an alternative to ReLU providing non-linearity to the model with faster convergence speed \cite{ElfwingUD18}. The attention weight is then used to compute the compositional embedding $\vh_{i}$ for item $v_i$:
\begin{equation}
    \vh_{i} = \sum_{n=1}^{N} \alpha_{n} \widetilde{\ve}_i^{\,n}.
    \label{eq:emb_aggregation}
\end{equation}
Finally, to facilitate side information modelling, we inject the context embedding of the item $i$ (i.e., $\vr_i$) into the above computed compositional embedding $h_i$ via a non-linear operation:
\begin{equation}
    \tilde{\vh}_i = \operatorname{MLP}([\vh_i;\vr_i]),
\end{equation}
where $[;]$ is the concatenation operation and $\operatorname{MLP}(\cdot):2D \rightarrow D$ denotes a multi-layer perceptron.
For a sequence of $T$ interacted items, we can obtain an embedding matrix $\mH = [\tilde{\vh}_{1};\tilde{\vh}_{2};...;\tilde{\vh}_{T}]^{\top}\in \mathbb{R}^{T\times D}$ by sequentially stacking all compositional item embeddings.

\subsection{Modelling Long- and Short-term User Preferences with Twin-Attention}
\label{TAN}
Self-attention has been a predominant approach in recent SRSs owing to its simplicity and capability of learning sequential dependencies among items. As discussed in Section \ref{sec:intro}, existing self-attentive sequential recommenders mostly deploy multiple attention heads in parallel to capture both global and local user preference signals from the sequence, resulting in unnecessary redundancy in both the network structure and parameter size. To alleviate such problem, we propose a twin-attention neural network to better capture the sequential information while maintaining the lightweight nature of LSAN. As depicted in Figure \ref{fig:overview}, it has two branches: a self-attention branch and a convolution branch specialised for global and local preference modelling, respectively.
\subsubsection{Convolution Branch for Local Patterns.}
Different from the self-attention branch which attends to all items in a sequence, convolution operations have shown success in extracting local features for image recognition and text classification. Their strong capacity in extracting regional features makes them an ideal component for capturing the short-term preferences among items that co-occur in a short time period. As such, with the matrix $\mH \in \sR^{T \times D}$ carrying all $T$ item embeddings in the sequence, we perform 1D convolution over the embedding matrix. Assuming the sliding window size is $L$ and the output size is $D$, the standard convolution operation needs $LD^2$ trainable parameters. To decrease the parameter size, we resort to a lightweight version of convolution \cite{WuFBDA19}. In particular, a depth-wise convolution operation is introduced, which applies a shared kernel of size $L$ for each channel (i.e., each item embedding dimension). This reduces the number of required parameters from $\mathcal{O}(LD^2)$ to $\mathcal{O}(LD)$. Mathematically, the $d$-th element ($d = 1,2,...,D$) of $i$-th embedding in the resulted output matrix $\mH^{conv} \in \sR^{T \times D}$ can be formulated as:
\begin{equation}
    \mH^{conv}_{i,d} = \sum_{j=1}^{L} \mW^{conv}_{j} \mH_{(i+j-\lceil\frac{L+1}{2}\rceil), d} \quad d=1,...,D,
\end{equation}
where $\mW^{conv}_{j} \in \sR^{L}$ is the kernel, and $\lceil \cdot \rceil$ denotes the ceiling operation. It is worth noting that, each row in the resulted matrix $\mH^{conv}$ encodes $i$-th item's interaction with items closely surrounding it within the $L$-sized sliding window, hence is a representation of all the local dependencies within the item sequence.  

\subsubsection{Self-attention Branch for Global Patterns.}
The rationale of coupling self-attention with convolution is that, by having a convolution branch dedicated to extracting local sequential patterns, the self-attention branch can now better specialise in learning global patterns, thus reducing the need for using an excessive amount of self-attention units for optimal performance.
As for self-attention, it has become one of the most prevalent means in various natural language processing (NLP) and sequential tasks as it can effectively capture relationships among items regardless of their distances (i.e., multi-hop). However, the plain self-attention fails to preserve the inherent orders of items in the sequence, impeding its efficacy for sequential recommendation tasks \cite{KangM18}. 
In this sense, we make the self-attention branch order-aware. Firstly, we define $T$ learnable position embeddings $\vp_1,\vp_2,...,\vp_T\in \sR^{D}$, which are stacked into a matrix $\mP \in \sR^{T \times D}$. Then, we fuse the positional information into the original item embeddings:
\begin{equation}
    \widetilde{\mH} = \mH + \mP,
\end{equation}
where each item is essentially paired with its corresponding positional context in the sequence. After that, a scaled dot-product self-attention is applied to compute item representations ${\mH}^{attn} \in \sR^{T\times D}$ by mining the long-range dependencies:
\begin{equation}
    \widehat{\mH} =\operatorname{softmax}(\frac{\mQ \mK^{\top}}{\sqrt{D / H}}) \mV,
\end{equation}
where $\mQ = \mW_q \widetilde{\mH}$, $\mK = \mW_k \widetilde{\mH}$ and $\mV = \mW_v \widetilde{\mH}$ are transformed item representations that are projected into query, key and value spaces, respectively. 

\subsubsection{Enhancing Expressiveness with Parallelism} Similar to pure self-attention-based methods, one can employ more than one attention heads for both branches in the twin-attention. For simplicity, we assume the convolution and self-attention modules each have $H$ heads in parallel. Then, the final output of the twin-attention can be obtained by concatenating $2H$ learned representation matrices followed by :
\begin{equation}
    \!\!\!\!\mH^{twin} = [{\mH}_1^{conv};...;{\mH}_H^{conv};{\mH}_1^{attn};...;{\mH}_H^{attn}],
\end{equation}
where $\mH^{twin}\in \sR^{T \times 2HD}$ is the final output. Note that in LSAN, it is not strictly necessary to set $H>1$ as the design of twin-attention can already facilitate comprehensively learning both global and local user preferences. One benefit of such parallelism over the traditional multi-head attention is that, with the same total amount of $2H$ attention heads, twin-attention consumes fewer parameters (i.e., $\mathcal{O}(H(LD + 3D^2))$ in twin attention versus $\mathcal{O}(6HD^2)$ in self-attention, $L\ll D$) and is able to yield stronger performance, as will be illustrated in Section \ref{sec:exp}.

\subsection{Prediction Layer}
\subsubsection{Point-wise Feed-forward Network}
To further enhance the 
 representation capacity of LSAN, we incorporate non-linearity into the output of the twin-attention. Specifically, we employ a point-wise feed-forward network (FFN) as follows:
\begin{equation}
    \widehat{\mH}^{twin} = \operatorname{GeLU}(\mH^{twin} \mW_{p}^{(1)} + \vb_p^{(1)}) \mW_{p}^{(2)} + \vb_p^{(2)},
\end{equation}
where $\mW_{p}^{(1)} \in \sR^{2HD \times 2HD}, \mW_{p}^{(2)} \in \sR^{2HD \times D}$ are weight matrices, $\vb_p^{(2)} \in \sR^{2H\times D},\vb_p^{(2)} \in \sR^{D}$ are bias vectors, and $\widehat{\mH}^{twin} \in sR^{T\times D}$ is the output of the point-wise FFN. Meanwhile, $\operatorname{GeLU}(\cdot)$ denotes the Gaussian error linear unit \cite{DevlinCLT19,HendrycksG16} that we use for non-linearity.


\subsubsection {Generating Rankings} With the final representation $\widehat{\mH}^{twin}$ that encodes both the user's long- and short-term interests, we generate the rankings for all items to facilitate top-$K$ recommendation. This is achieved by estimating the likelihood of having user $u$ interact with each item, which is formulated as learning a $|\mathcal{V}|$-dimensional probability distribution $\widehat{\textbf{y}}_t$:
\begin{equation}
    \widehat{\textbf{y}} = \operatorname{softmax}(\mW_{o} \widehat{\mH}^{twin} + \vb_o),
\end{equation}
where $\mW_{o} \in \sR^{|\gV| \times D}$ and $\vb_o \in \sR^{|\gV|}$ are the learnable weight matrix and bias vector, respectively. By sorting each $v_i$ according to its corresponding probability score $\widehat{y}_i \in \widehat{\textbf{y}}$ in a descending order, we will be able to truncate $K$ items from the top of the list as our recommendation results. 

\subsection{Learning Objective}
With the estimated probability vector $\widehat{\textbf{y}}$, we then employ cross-entropy loss function to quantify the error of predicting the next item for LSAN:
\begin{equation}
    \mathcal{L} = - \frac{1}{S} \sum_{s=1}^{S} \textbf{y}^{\top}_s \log(\widehat{\textbf{y}}_s) + \lambda\|\Psi\|^{2}_{2},
\end{equation}
where $s\leq S$ is the index of training samples, $\textbf{y}_s$ is the one-hot vector representing the ground truth of the next item, and $\Psi$ is the set of all trainable parameters under the $L2$ regularisation term with coefficient $\lambda$. 

\section{Experiments}\label{sec:exp}
In this section, we evaluate the recommendation effectiveness and memory efficiency of our LSAN model for sequential recommendation. Specifically, we first analyze the performance of LSAN by comparing it with state-of-the-art sequential recommenders from both accuracy and model size perspectives. After that, we further investigate the impact of the key components and hyperparameters in LSAN. 

\begin{table}[!htb]
\centering
\caption{Statistics of experimental datasets.}
\begin{tabular}{ccccccc}
\hline
Datasets & Beauty  & Toys    & Sports  & Yelp    \\
\hline
\#Users              & 22,363  & 19,412   & 35,598  & 22,845   \\
\#Items              & 12,101  & 11,924   & 18,357  & 16,552   \\
\#Categories         & 6       & 24      & 35      & 22       \\
\#Interactions          & 198,502 & 167,597  & 296,337 & 243,703  \\
Avg. Int. per User & 8.8764  & 8.6337  & 8.3245  & 10.6677 \\
Avg. Int. per Item & 16.4038 & 14.0554 & 16.1430 & 14.7235 \\
Sparsity             & 99.93\% & 99.93\% & 99.95\% & 99.94\% \\
\hline
\end{tabular}
\label{tab:dataset_statistics}
\end{table}

\subsection{Experimental Settings}
\subsubsection{Dataset}
We conduct experiments on four commonly-used benchmark datasets. The statistical details of all datasets after preprocessing are reported in Table \ref{tab:dataset_statistics}, including the number of users, items, interactions, categories, average interactions per user (Avg. Int./User) and average interactions per item (Avg. Int./Item). All the experimental datasets are highly sparse. We briefly introduce their properties below.
\begin{itemize}
    \item Beauty, Sports and Toys\footnote{http://jmcauley.ucsd.edu/data/amazon/links.html}: These three datasets are provided by \cite{HeM16}, which are collected from Amazon and contain product reviews and abundant metadata. 
    \item Yelp\footnote{https://www.yelp.com/dataset}: The dataset contains user check-in data provided by Yelp, where businesses are viewed as items. The data we use for our experiments span across 2019.
\end{itemize}
For each dataset, we group interactions by user IDs, and then generate one chronological item sequence for each user. The inactive users and unpopular items with less than 5 interactions are discarded.

\subsection{Evaluation Metrics}
We adopt the \textit{leave-one-out} evaluation approach, i.e., for each user interaction sequence, we use the last item as the test instance, the second last item as a validation sample, and the remaining items for training. We choose Hit Ratio at Rank $K$ (HR$@K$) and Normalised Discounted Cumulative Gain at Rank $K$ (nDCG$@K$) on top-$K$ ranked items, which are widely used in recommender systems \cite{chen2019air,KangM18} for top-$K$ performance and overall ranking performance evaluation. We report the performance results on HR$@$\{5, 10, 20\} and nDCG$@$\{5, 10, 20\}, respectively. As suggested by \cite{KricheneR20}, to eliminate potential biases, we rank each ground truth item along with the whole item set (i.e., $\gV$) to compute all metrics, and report the average scores over all users. 

\subsection{Baseline Methods}
We compare LSAN with the most representative, state-of-the-art sequential recommendation methods below:
\begin{itemize}
	\item FPMC \cite{RendleFS10}: It is a combination of matrix factorisation with Markov chain, which can simultaneously capture sequential information and long-term user preferences.
	\item GRU4Rec \cite{HidasiKBT15}: It is an RNN-based sequential recommender with session-wise mini-batch training strategy. The model is optimised by a pair-wise ranking loss.
	\item Caser \cite{TangW18}: This is a CNN-based method that models high-order Markov-chain probability by performing convolutional operations on the item embedding matrix.
	\item SASRec \cite{KangM18}: It is a next-item sequential recommendation method based on the Transformer architecture, which employs multi-head self-attention mechanism to explore implicit user interactions. 
	\item BERT4Rec \cite{SunLWPLOJ19}: It is an improvement of SASRec, which contains an additional Cloze objective and bidirectional self-attention structure.
\end{itemize}

\subsection{Implementation Details}
LSAN is implemented using PyTorch with Nvidia GTX 2080 Ti. In LSAN, we set the dimension size $D$ to 128, CNN kernel size $L$ to 5, the number of attention heads $H$ to 2 for each branch, and the number of stacked twin-attention layers to 1 on all datasets. All the trainable parameters in our model are optimised using Adam optimiser \cite{KingmaB14} with the batch size of 256, learning rate of 0.001 and $L2$ regularisation strength $\lambda$ of $1e-5$. For a fair comparison on accuracy and model size, we apply the same dimension size $D$ for all methods' embeddings. Note that in LSAN, altering either $N$ or $m_n$ can lead to different compression rates on the original embedding table. Hence, we fix $N=2$ and vary $m_1$ ($m_2 = \frac{|\mathcal{V}|}{m_1}$) for the ease of hyper-parameter tuning. In Section \ref{sec:opc}, we will first test LSAN's performance with $m_1=2$, while we will further discuss how LSAN performs when we compress the model size more aggressively with a larger $m_1$ in Section \ref{sec:comp_rate}.
 
\begin{table*}[!htb]
\caption{Comparison on sequential recommendation accuracy and model sizes. In each row, the best and second best results are highlighted in boldface and underlined, respectively. The parameter size of each model is obtained when $D=128$.}
\begin{tabular}{clccccccrcr}
\hline
Datasets                & \multicolumn{1}{c}{Metrics}  & FPMC   & GRU4Rec & Caser  & SASRec  & BERT4Rec      & \textbf{LSAN}$_{full.emb}$      & Improv.  & \textbf{LSAN}            & Improv.  \\
\hline
\hline
\multirow{7}{*}{Beauty} & HR$@$5            & 0.0149        & 0.0164           & 0.0205         & {\ul 0.0419}    & 0.0312    &        0.0432          & 3.10\%  & \textbf{0.0492} & 17.42\% \\
                                 & HR$@$10           & 0.0273        & 0.0283           & 0.0347         & {\ul 0.0650}    & 0.0468            & 0.067           & 3.08\%  & \textbf{0.0785} & 20.77\% \\
                                 & HR$@$20           & 0.0438        & 0.0479           & 0.0556         & {\ul 0.0872}    & 0.0737            & 0.0992          & 13.76\% & \textbf{0.1201} & 37.73\% \\
                                 & nDCG$@$5          & 0.0096        & 0.0099           & 0.0131         & {\ul 0.0263}    & 0.0223            & 0.0276          & 4.94\%  & \textbf{0.0316} & 20.15\% \\
                                 & nDCG$@$10         & 0.0133        & 0.0137           & 0.0176         & {\ul 0.0337}    & 0.0272            & 0.0352          & 4.45\%  & \textbf{0.041}  & 21.66\% \\
                                 & nDCG$@$20         & 0.0173        & 0.0187           & 0.0229         & {\ul 0.0372}    & 0.0340            & 0.0433          & 16.4\%  & \textbf{0.0515} & 38.44\% \\
                                 & \#Parameters    & 8.26M   & 4.06M  & 8.42M  & 1.75M   & 4.29M     & 1.71M       &  \multicolumn{1}{c}{-}       & 1.11M   & \multicolumn{1}{c}{-}        \\
\hline
\multirow{7}{*}{Toys}   & HR$@$5            & 0.0099        & 0.0097           & 0.0166         & {\ul 0.0450}    & 0.0136            & 0.045           & 0.00\%  & 0.0437          & -2.89\% \\
                                 & HR$@$10           & 0.0175        & 0.0176           & 0.0270         & {\ul 0.0650}    & 0.0195            & 0.0676          & 4.00\%  & \textbf{0.0711} & 9.38\%  \\
                                 & HR$@$20           & 0.0273        & 0.0301           & 0.0420         & {\ul 0.0925}    & 0.0333            & 0.097           & 4.86\%  & \textbf{0.1181} & 27.68\% \\
                                 & nDCG$@$5          & 0.0064        & 0.0059           & 0.0107         & {\ul 0.0300}    & 0.0077            & \textbf{0.0305} & 1.67\%  & 0.0283          & -5.67\% \\
                                 & nDCG$@$10         & 0.0088        & 0.0084           & 0.0141         & {\ul 0.0370}    & 0.0096            & \textbf{0.0378} & 2.16\%  & 0.037           & 0.00\%     \\
                                 & nDCG$@$20         & 0.0112        & 0.0116           & 0.0179         & {\ul 0.0436}    & 0.0130            & 0.0452          & 3.67\%  & \textbf{0.0488} & 11.93\% \\
                                 & \#Parameters    & 7.77M  & 4.01M   & 7.9M   & 1.73M   & 4.24M   & 1.68M       & \multicolumn{1}{c}{-}       & 1.28M   & \multicolumn{1}{c}{-}        \\
\hline
\multirow{7}{*}{Sports} & HR$@$5            & 0.0088        & 0.0129           & 0.0116         & {\ul 0.0201}    & 0.0139            & 0.0229          & 13.93\% & \textbf{0.0314} & 56.22\% \\
                                 & HR$@$10           & 0.0160        & 0.0204           & 0.0194         & {\ul 0.0314}    & 0.0207            & 0.0366          & 16.56\% & \textbf{0.0481} & 53.18\% \\
                                 & HR$@$20           & 0.0259        & 0.0333           & 0.0314         & {\ul 0.0486}    & 0.0438            & 0.0578          & 18.93\% & \textbf{0.0759} & 56.17\% \\
                                 & nDCG$@$5          & 0.0055        & 0.0086           & 0.0072         & {\ul 0.0129}    & 0.0085            & 0.0146          & 13.18\% & \textbf{0.0211} & 63.57\% \\
                                 & nDCG$@$10         & 0.0077        & 0.0110           & 0.0097         & {\ul 0.0164}    & 0.0106            & 0.0191          & 16.46\% & \textbf{0.0264} & 60.98\% \\
                                 & nDCG$@$20         & 0.0100        & 0.0142           & 0.0126         & {\ul 0.0208}    & 0.0162            & 0.0244          & 17.31\% & \textbf{0.0334} & 60.58\% \\
                                 & \#Parameters    & 12.76M & 5.83M    & 12.93M & 2.55M   & 6.05M   & 2.51M       & \multicolumn{1}{c}{-}       & 1.73M   & \multicolumn{1}{c}{-}        \\
\hline
\multirow{7}{*}{Yelp}   & HR$@$5            & 0.0116        & 0.0152           & 0.0151         & {\ul 0.0210}    & 0.0184            & 0.0251          & 19.52\% & \textbf{0.0385} & 83.33\% \\
                                 & HR$@$10           & 0.0211        & 0.0263           & 0.0253         & {\ul 0.0356}    & 0.0259            & 0.0451          & 26.69\% & \textbf{0.0682} & 91.57\% \\
                                 & HR$@$20           & 0.0352        & 0.0439           & 0.0422         & {\ul 0.0575}    & 0.0430            & 0.0744          & 29.39\% & \textbf{0.1148} & 99.65\% \\
                                 & nDCG$@$5          & 0.0074        & 0.0099           & 0.0096         & {\ul 0.0126}    & 0.0114            & 0.0157          & 24.6\%  & \textbf{0.0205} & 62.7\%  \\
                                 & nDCG$@$10         & 0.0103        & 0.0134           & 0.0129         & {\ul 0.0176}    & 0.0138            & 0.0221          & 25.57\% & \textbf{0.0301} & 71.02\% \\
                                 & nDCG$@$20         & 0.0137        & 0.0178           & 0.0171         & {\ul 0.0230}    & 0.0181            & 0.0294          & 27.83\% & \textbf{0.0417} & 81.3\%  \\
                                 & \#Parameters    & 10.20M  & 5.32M   & 10.37M  & 2.32M   & 5.61M     & 2.27M       & \multicolumn{1}{c}{-}       & 1.53M   & \multicolumn{1}{c}{-} \\
\hline
\end{tabular}
\label{tab:model_performance}
\end{table*}

\subsection{Overall Performance Comparison}\label{sec:opc}
We summarise the results of all models on four benchmark datasets in Table \ref{tab:model_performance}. From the table, we can draw the following observations:

Among all sequential baseline methods, FPMC receives the worst results over all evaluation metrics. This is mainly because FPMC only exploits first-order dependencies where the higher-order relationships among items are neglected. In comparison, Caser utilises convolutional kernels to extract k-hop adjacent item dependencies, thus, obtaining better performance results than FPMC. However, since the sliding window size of Caser could only cover a small number of items, which lacks the ability on handling sparse datasets with long sequences, e.g., Yelp. As a result, it can be observed that the RNN-based method, GRU4Rec, has better performance than Caser on those datasets. SASRec and BERT4Rec are state-of-the-art self-attentive approaches, which have clear margins with the other baseline methods. This shows the superiority of the self-attention architecture in sequential behaviour modelling. It is worth noting that BERT4Rec does not outperform SASRec under our evaluation settings (i.e., full test sample set), which is different from their original paper. We think this maybe because there may exist improper biases in the naive negative sampled test sets according to their original implementation details.

We present the performance results and relative improvements over the best baseline of two versions of our proposed LSAN in Table \ref{tab:model_performance}, i.e., LSAN$_{full.emb}$ and LSAN. The only difference between them is that LSAN is our proposed model, while LSAN$_{full.emb}$ is trained using a full-sized embedding table. From the results, using either full-sized embedding table or our proposed dynamic context-aware compositional embedding can surpass the best baseline method SASRec with significant margins on most evaluation metrics. This proves the effectiveness of our twin-attention structured model on handling long-term and short-term user preferences. Moreover, though a large number of parameters are reduced via our compositional embeddings, LSAN performs even much better than LSAN$_{full.emb}$ on all datasets. We believe that by incorporating seasonal and categorical factors (i.e., temporal dynamics) within the model can largely enhance the expressiveness of LSAN with fewer trainable parameters. 

In addition, an obvious model size reduction can be observed from both LSAN and LSAN$_{full.emb}$. We provide a more detailed discussion on memory usage in the following section.

\begin{table}[!htb]
\caption{A comparison of performance results and number of model parameters using different embedding compression rate $m_1$ on four datasets.}
\renewcommand{\arraystretch}{1.1}
\setlength\tabcolsep{2pt}
\resizebox{0.49\textwidth}{!}{
\begin{tabular}{llccccc}
\hline
Datasets       & \multicolumn{1}{c}{Metrics} & SASRec & LSAN(2x) & LSAN(3x) & LSAN(4x) & LSAN(5x) \\
\hline
\multirow{4}{*}{Beauty} & HR$@$20           & 0.0872          & 0.1201            & 0.0981            & 0.043             & 0.0456            \\
                        & nDCG$@$20         & 0.0372          & 0.0515             & 0.0385            & 0.0158            & 0.0178            \\
                        & \#Parameters    & 1.75M            & 1.11M              & 0.71M              & 0.58M             & 0.5M               \\
                        & Relative  Size  & 100.00\%        & 63.43\%           & 40.57\%           & 33.14\%           & 28.57\%           \\
\hline
\multirow{4}{*}{Toys}   & HR$@$20           & 0.0925          & 0.1181            & 0.0887            & 0.0618            & 0.0539            \\
                        & nDCG$@$20         & 0.0436          & 0.0488            & 0.0341            & 0.0232            & 0.0211            \\
                        & \#Parameters    & 1.73M            & 1.28M              & 0.88M              & 0.75M              & 0.68M              \\
                        & Relative  Size  & 100.00\%        & 73.99\%           & 50.87\%           & 43.35\%           & 39.31\%           \\
\hline
\multirow{4}{*}{Sports} & HR$@$20           & 0.0486          & 0.0759            & 0.0551            & 0.0370            & 0.0311            \\
                        & nDCG$@$20         & 0.0208          & 0.0334            & 0.0249            & 0.0159             & 0.0125            \\
                        & \#Parameters    & 2.55M            & 1.73M              & 1.34M              & 1.14M                 & 1.03M              \\
                        & Relative  Size  & 100.00\%        & 67.84\%           & 52.55\%           & 44.71\%           & 40.39\%           \\
\hline
\multirow{4}{*}{Yelp}   & HR$@$20           & 0.0575          & 0.1148            & 0.1087            & 0.0472            & 0.0436            \\
                        & nDCG$@$20         & 0.023           & 0.0417            & 0.0434            & 0.0187            & 0.0161            \\
                        & \#Parameters    & 2.32M            & 1.53M              & 1.03M              & 0.85M              & 0.74M              \\
                        & Relative  Size  & 100.00\%        & 65.95\%           & 44.40\%           & 36.64\%           & 31.90\%   \\
\hline
\end{tabular}}
\label{tab:compression_rate}
\end{table}

\subsection{Impact of Embedding Compression Rate}\label{sec:comp_rate}
In this section, we study how the compression rate affects the model performance. Intuitively, the model will have worse performance when the compression rate $m_1$ increases. We compare the model size and next-item recommendation performance of LSAN at different compression rates with the best baseline method, SASRec. From the results shown in Table \ref{tab:compression_rate}, we can observe that when $m_1=2$, our LSAN model surpasses SASRec over most metrics with only around 60\% of parameters in SASRec on all datasets. When $m_1$ increases to 3 (i.e., around 45\% of SASRec parameters), LSAN still produce comparable recommendation results on Beauty, Toys and Yelp datasets, which further approves the outstanding memory efficiency and recommendation effectiveness of our model. However, when $m_1$ goes up greater than 3, we can see a clear performance drop. It is understandable that when the model contains a very small number of parameters, the model is under-parameterised resulting in the failure of capturing meaningful information from items.

\begin{figure*}
    \centering
    \includegraphics[height=5cm]{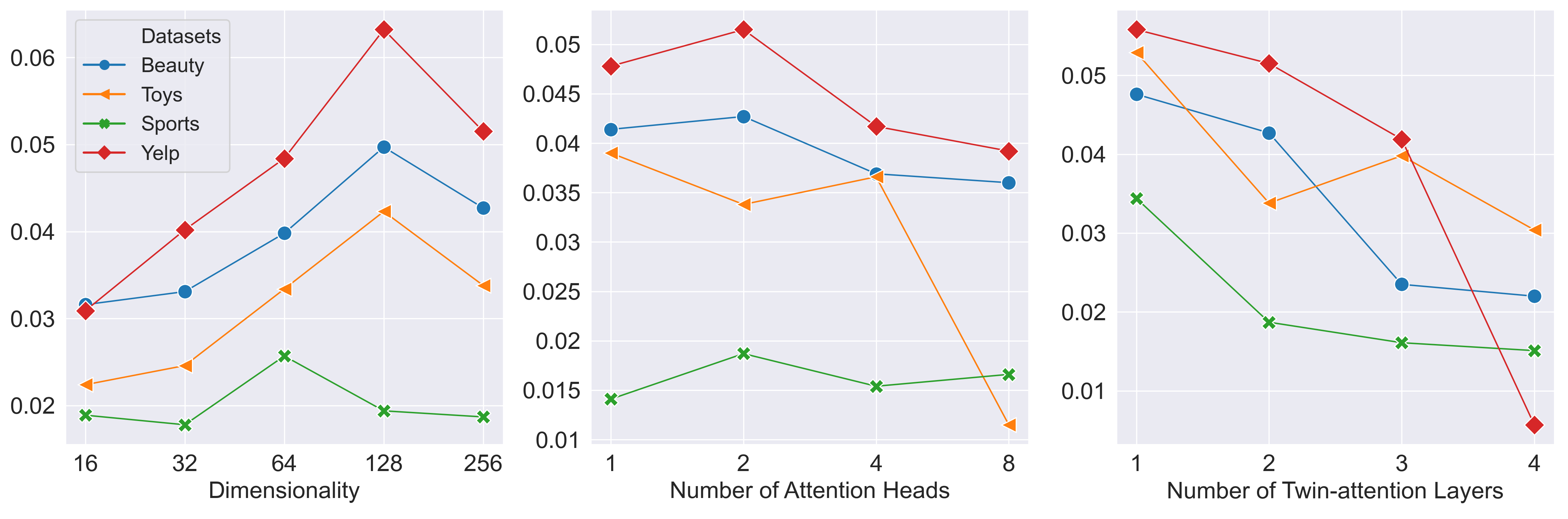}
    \vspace{-0.4cm}
    \caption{Effect of dimensionality, number of attention heads, number of twin-attention layers (nDCG$@$20).}
    \label{fig:hyperparameter}
\end{figure*}

\begin{table}[!htb]
\caption{Ablation study of different variants on four datasets.}
\resizebox{0.48\textwidth}{!}{
\begin{tabular}{llccc}
\hline
\multirow{2}{*}{Datasets} & \multicolumn{4}{c}{Variants}                 \\
                          \cline{2-5}
                          & Metrics & LSAN$_{w/o.dynamic}$ & LSAN$_{plain.attn}$ & LSAN  \\
                          \hline
\multirow{2}{*}{Beauty}   & HR$@$20   & 0.977   & 0.108            & 0.120 \\
                          & nDCG$@$20 & 0.038   & 0.045            & 0.051 \\
\hline
\multirow{2}{*}{Toys}     & HR$@$20   & 0.094   & 0.063            & 0.118 \\
                          & nDCG$@$20 & 0.033   & 0.021            & 0.049 \\
\hline
\multirow{2}{*}{Sports}   & HR$@$20   & 0.076   & 0.066            & 0.076 \\
                          & nDCG$@$20 & 0.033   & 0.030            & 0.033 \\
\hline
\multirow{2}{*}{Yelp}     & HR$@$20   & 0.104   & 0.011            & 0.115 \\
                          & nDCG$@$20 & 0.038   & 0.046            & 0.042 \\
\hline
\end{tabular}}
\label{tab:ablation_study}
\end{table}

\subsection{Ablation Study}
Comparing with the existing self-attentive methods, LSAN mainly contains two novel components: dynamic context-aware compositional embeddings and twin-attention layers. To verify the effectiveness of each component, we conduct ablation studies on all benchmark datasets. Table \ref{tab:ablation_study} shows the performance of our default model and its two degraded variants ($D=128$). We give a brief description and detailed analysis of each variant in the following:
Specifically, we compare LSAN with the following degraded variants:
\begin{itemize}
    \item LSAN$_{w/o.dynamic}$: This variant removes the dynamic compositional embedding component. The embedding part becomes exactly the same as the QR embedding introduced in \cite{ShiMNY20}. More concretely, we create this variant by modifying Eq. (\ref{eq:emb_aggregation}) to $\vh_{i} = \sum_{n=1}^{N}\widetilde{\ve}_i^{\,n}$. We can see a significant performance drop on most datasets when the temporal information is removed from the composited embeddings. This suggests that respecting temporal dynamics is of great importance in user preference modelling.
    \item LSAN$_{plain.attn}$: This variant replaces the twin-attention layers with self-attention layers. There is a clear performance drop on three datasets when only self-attention is applied. This reveals that the self-attention does not have sufficient capability to uncover both long-term and short-term user preferences with limited attention heads.
\end{itemize}

\subsection{Hyper-parameter Analysis}
\label{sec:hyperparameter}
We further examine the impact of four various hyper-parameters, including dimension size $D$, partition size $m$, number of attention heads $H$, and number of stacked twin-attention layers. For each test, we vary the value of one hyper-parameter, while keep the others be the optimal settings. The results are demonstrated in Figure \ref{fig:hyperparameter}.

\subsubsection{Impact of Dimension Size.} the dimension of
The value of dimension size is examined from 16 to 256. We can observe that a small dimension size (i.e., 16) cannot preserve sufficient latent information of items for user sequential behaviour modelling. The model performance increases steadily when the dimension size grows up on all datasets. However, we also find that a larger hidden dimensionality (e.g., 256) may not contribute to better performance, which may be mainly caused by the over-fitting problem especially when the data is extremely sparse. This also proved by the performance results on Sports dataset that has highest sparsity among all datasets: LSAN becomes sensitive to the choice of appropriate dimensionality, i.e., $D=64$ fits the model best. On the other datasets, LSAN reaches the best performance when $D=128$.

\subsubsection{Impact of Attention Head Number.}
The results in the second graph in Figure \ref{fig:hyperparameter} show that more attention heads contribute better to the model performance. It is worth noting that the actual number of our twin-attention heads are doubled. Thus, LSAN model reaches best performance on most datasets, when it is equipped with 2 self-attention heads and 2 convolution heads.

\subsubsection{Impact of Twin-attention Layer Number.}
LSAN receives the best performance with 1-layer architecture. Different from SASRec and BERT4Rec, which usually require 2 or 3 layers to fully capture various-order item dependencies. With the help of our proposed twin-attention structure, our model is capable of capturing various sequential information with only one layer. We also observe that the model performance drops obviously when more twin-attention layers are stacked on all dataset. This may be because of the over-fitting problem when LSAN is launched on extremely sparse datasets.

\begin{figure}[t]
 \centering
 \begin{minipage}{0.49\columnwidth}
 \subfloat[][]
     {\includegraphics[width=\columnwidth]{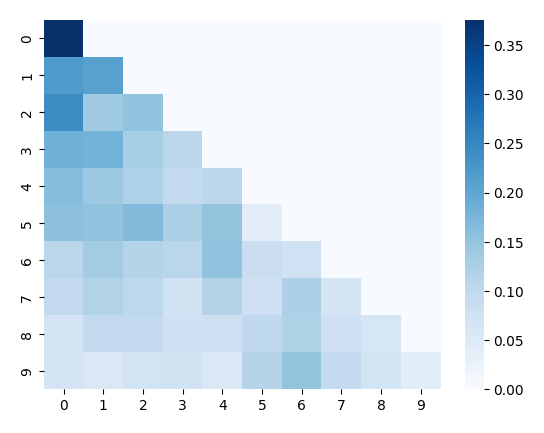}} \\
     \subfloat[][]
     {\includegraphics[width=\columnwidth]{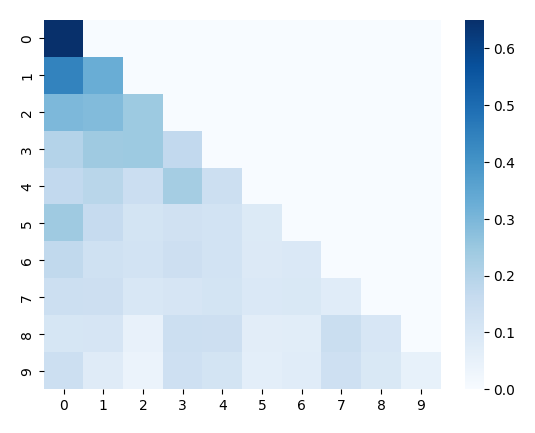}} \\
     \subfloat[][]
     {\includegraphics[width=\columnwidth]{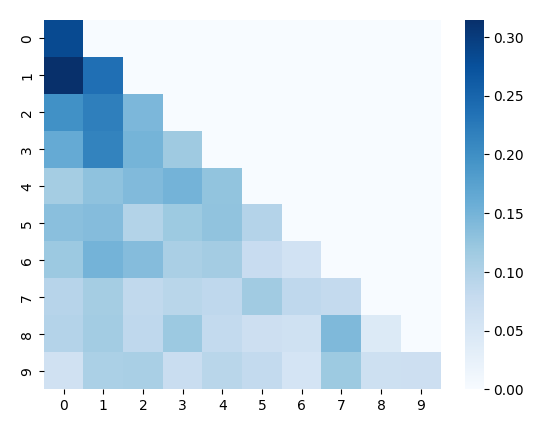}}
 \end{minipage}
 \begin{minipage}{0.49\columnwidth}
    \subfloat[][]
     {\includegraphics[width=\columnwidth]{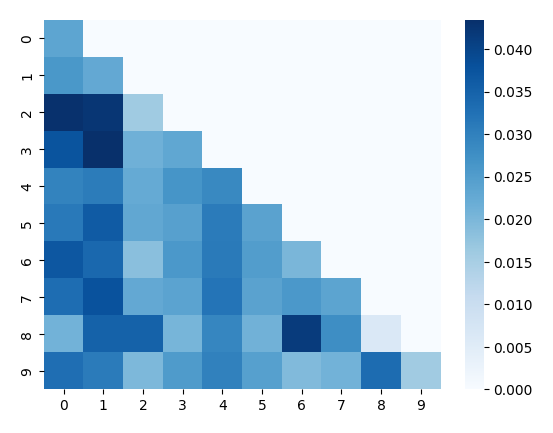}} \\
     \subfloat[][]
     {\includegraphics[width=\columnwidth]{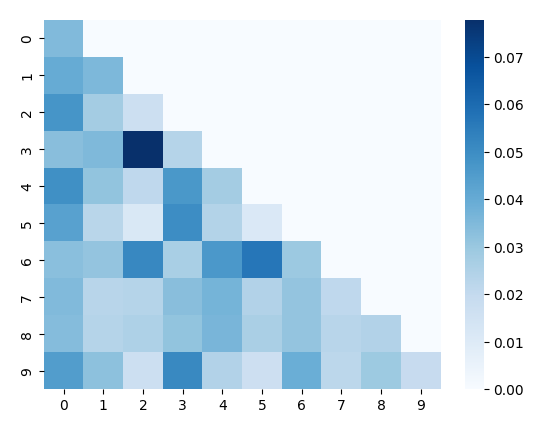}} \\
     \subfloat[][]
     {\includegraphics[width=\columnwidth]{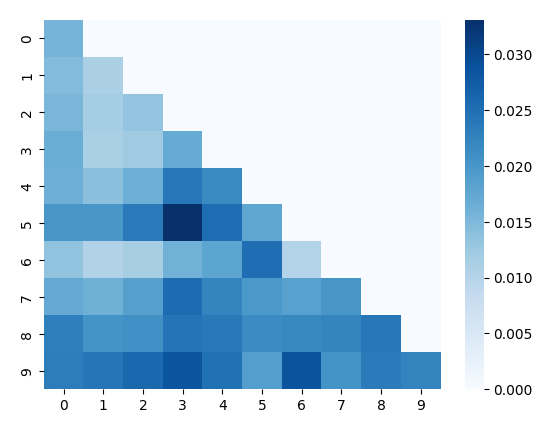}}
 \end{minipage}
 \caption{Heat maps of average attention weights on Yelp dataset. (a-c) are from SASRec, while (d-f) are from LSAN's self-attention branch.}
 \label{fig:heatmap}
\end{figure}

\subsection{Attention Weight Visualisation}
Recall that in Section \ref{sec:intro}, we argue that self-attention-based models put too much emphasise on local patterns, which is the motivation of our design on twin-attention architecture. To more intuitively demonstrate how twin-attention performs effective user behaviour modelling, we examine three randomly selected user interaction sequence samples and calculate the averaged attention weights on the last 10 items from each sample over all attention heads. The heat maps of the normalised attention weights from LASN and SASRec are illustrated in Figure \ref{fig:heatmap}. From the figure, it can be easily distinguished that the attention module in LSAN focus on item global patterns (i.e., no diagonal pattern shown in the figure), thus leaving local pattern modelling to the convolution branch in our twin attention. In comparison, the heat maps of SASRec show a clear concentration on local patterns, which fails to model global patterns effectively.

\section{Related Work}
\subsection{Sequential Recommendation}
Early work on sequential recommendation is mainly based on Markov chains. Rendle et al. \cite{RendleFS10} propose FPMC that combines the power of matrix factorisation and Markov-chain to learn an item-to-item transition probability matrix, which is then used to make next item prediction based on the user's latest interaction. After that, several models built upon high-order Markov chains are introduced \cite{HeM16,HeKM17}. The advances in recurrent neural networks (RNNs) have brought significant performance boost in sequential recommendation. Hidasi et al. \cite{HidasiKBT15} propose a sequential recommender based on RNNs, which employs gated recurrent units (GRUs) to extract the high-order sequential information from the user's interaction history. Subsequent RNN-based approaches leverage attention networks \cite{SunQCLNY20}, memory networks \cite{HuangRZHWD19,HuangZDWC18}, copy mechanism \cite{RenCLR0R19}, or reinforcement learning scheme \cite{XinKAJ20}, to improve the effectiveness of sequential user interest modelling. Another line of work \cite{TangW18} treats a sequence of item embeddings as a feature map of an image, and performs convolution operation upon embeddings to capture local dependencies among items.

Owing to the promising capability in sequential data modelling, attention mechanism has become popular and widely studied in various domains, such as text classification \cite{YangYDHSH16} and machine translation \cite{BahdanauCB14}. However, these approaches treat attention mechanism as an additional module upon the RNN backbone, resulting in higher computational cost. To solve this issue, a new attention architecture, transformer, is proposed in \cite{VaswaniSPUJGKP17,DevlinCLT19}. Its main building block is multi-head self-attention, which allows faster parallel computation and achieves state-of-the-art performance in a wide range of sequence modelling tasks. In light of self-attention, Kang et al. \cite{KangM18} propose a self-attentive framework named SASRec, which adopts a multi-head self-attention layer to capture the user's sequential behaviours and achieves state-of-the-art performance on various sequential datasets. Later on, Sun et al. \cite{SunLWPLOJ19} encode sequence data in bidirectional manner by introducing BERT4Rec together with a masked training scheme. However, all aforementioned sequential recommendation methods suffer from high memory cost in two aspects, which are infeasible for on-device applications. First, the large item embedding table brings high memory complexity. Second, as discussed in \cite{WuLLLH20}, the multi-head self-attention architecture tends to pay more attention on local dependencies, resulting in weak global preference modelling. In contrast, our proposed LSAN largely reduces the memory cost from the embedding table by a dynamic compositional embedding scheme. Besides, LSAN effectively learns global and local dependencies by a novel twin-attention, where the heavy redundancy in traditional self-attentive recommenders are resolved by two specialised branches for long- and short-range pattern mining.

\subsection{Lightweight Deep Learning Models}
DNN-based methods have demonstrated strong capability in various recommendation tasks. However, with the rapid development of edge devices, there has been an increasing demand on adopting DNN-based models on mobile phones and even smaller edge devices for stability and reliability. In this line of research, the methodologies can be roughly categorised into four types: pruning, quantisation, knowledge distillation, and compositional embedding. Network pruning approaches manage to reduce the over-fitting parameters by discarding unnecessary ones from the neural model. Zhou et al. \cite{ZhouAP16} introduce a group-sparse regularisation upon CNN kernel to produce a compact version without losing accuracy. However, most pruning methods require more iterations to reach convergence leading to extreme time cost. The second line of work aim to create one or more codebooks for a group of similar item representations. J{\'{e}}gou et al. \cite{JegouDS11} propose to decompose the item representation space into multiple subspaces, and then a codebook of each subspace can be obtained by clustering items in each subspace. The recent work, LightRec \cite{LianWLLC020}, develops a recurrent composite embedding encoder to learn diversified codebooks in a recursive manner. However, the learning of codebooks is independent of learning the item representations. Thus, the model cannot be trained end-to-end. Recently, knowledge distillation has gained popularity due to its high adaptivity to various complex models. This line of work primarily trains a large complex teacher network at the first place, and subsequently utilises soft labels obtained from the teacher model to train a lightweight student model. Wang et al. \cite{WangYCHWZH20} devise a tensor-train decomposed lightweight RNN model, and train the model using a well-tuned state-of-the-art teacher model via knowledge distillation for next POI recommendation. With the observation that the embedding matrices in recommender systems are the major source of memory consumption, some recent studies resort to embedding compression. A number of studies \cite{ShuN18,ChenMS18,ZhangYHDYL18,ZhangXLLH19,YadanLowBit20,ZhangHash21} introduce the idea of converting a continuous-valued embedding vector to a discrete code, where each bit refers to the learned index of a base embedding table. However, this still requires the model to store extra discrete code for each item. To solve the limitation, Shi et al. \cite{ShiMNY20} propose a quotient-remainder indexing technique, which is able to obtain a unique set of base embeddings without allocating extra embedding space. Nevertheless, all the above-mentioned embedding compression work is designed for static recommendation scenario, which neglects the dynamics of user interests in sequential recommendation. To address this issue, we design a context-aware temporal compositional embedding scheme that incorporates temporal information by attentively merging base embeddings for each item. As such, our proposed LSAN is capable of preserving the temporal dynamics and optimising memory efficiency simultaneously.

\section{Conclusion}
In this paper, we introduce a lightweight twin-attention sequential recommender named LSAN, where two parallel branches are respectively specialised for short-term and long-term user preference modelling. To overcome the common bottleneck of large memory cost in existing DNN-based sequential recommender, we introduce temporal context-aware compositional embedding scheme, which largely reduces the memory cost and preserves intrinsic temporal dynamics of sequential data. Extensive experiments conducted on four real-world datasets clearly demonstrate the effectiveness and efficiency of our proposed model.

\begin{acks}
This work is partially supported by the Australian Research Council under the streams of Discovery Project (No. DP190101985), Future Fellowship (No. FT210100624), Centre of Excellence (No. CE200100025), and Industry Transformation Training Centre (No. IC200100022).
\end{acks}

\bibliographystyle{ACM-Reference-Format}
\bibliography{cikm21}


\begin{thebibliography}{55}


\ifx \showCODEN    \undefined \def \showCODEN     #1{\unskip}     \fi
\ifx \showDOI      \undefined \def \showDOI       #1{#1}\fi
\ifx \showISBNx    \undefined \def \showISBNx     #1{\unskip}     \fi
\ifx \showISBNxiii \undefined \def \showISBNxiii  #1{\unskip}     \fi
\ifx \showISSN     \undefined \def \showISSN      #1{\unskip}     \fi
\ifx \showLCCN     \undefined \def \showLCCN      #1{\unskip}     \fi
\ifx \shownote     \undefined \def \shownote      #1{#1}          \fi
\ifx \showarticletitle \undefined \def \showarticletitle #1{#1}   \fi
\ifx \showURL      \undefined \def \showURL       {\relax}        \fi
\providecommand\bibfield[2]{#2}
\providecommand\bibinfo[2]{#2}
\providecommand\natexlab[1]{#1}
\providecommand\showeprint[2][]{arXiv:#2}

\bibitem[\protect\citeauthoryear{Bahdanau, Cho, and Bengio}{Bahdanau
  et~al\mbox{.}}{2015}]%
        {BahdanauCB14}
\bibfield{author}{\bibinfo{person}{Dzmitry Bahdanau},
  \bibinfo{person}{Kyunghyun Cho}, {and} \bibinfo{person}{Yoshua Bengio}.}
  \bibinfo{year}{2015}\natexlab{}.
\newblock \showarticletitle{Neural Machine Translation by Jointly Learning to
  Align and Translate}. In \bibinfo{booktitle}{\emph{3rd International
  Conference on Learning Representations, {ICLR} 2015, San Diego, CA, USA, May
  7-9, 2015, Conference Track Proceedings}},
  \bibfield{editor}{\bibinfo{person}{Yoshua Bengio} {and} \bibinfo{person}{Yann
  LeCun}} (Eds.).
\newblock


\bibitem[\protect\citeauthoryear{Chen, Min, and Sun}{Chen
  et~al\mbox{.}}{2018a}]%
        {ChenMS18}
\bibfield{author}{\bibinfo{person}{Ting Chen}, \bibinfo{person}{Martin~Renqiang
  Min}, {and} \bibinfo{person}{Yizhou Sun}.} \bibinfo{year}{2018}\natexlab{a}.
\newblock \showarticletitle{Learning K-way D-dimensional Discrete Codes for
  Compact Embedding Representations}. In \bibinfo{booktitle}{\emph{Proceedings
  of the 35th International Conference on Machine Learning, {ICML} 2018,
  Stockholmsm{\"{a}}ssan, Stockholm, Sweden, July 10-15, 2018}}
  \emph{(\bibinfo{series}{Proceedings of Machine Learning Research},
  Vol.~\bibinfo{volume}{80})}. \bibinfo{publisher}{{PMLR}},
  \bibinfo{pages}{853--862}.
\newblock


\bibitem[\protect\citeauthoryear{Chen, Yin, Chen, Yan, Nguyen, and Li}{Chen
  et~al\mbox{.}}{2019}]%
        {chen2019air}
\bibfield{author}{\bibinfo{person}{Tong Chen}, \bibinfo{person}{Hongzhi Yin},
  \bibinfo{person}{Hongxu Chen}, \bibinfo{person}{Rui Yan},
  \bibinfo{person}{Quoc Viet~Hung Nguyen}, {and} \bibinfo{person}{Xue Li}.}
  \bibinfo{year}{2019}\natexlab{}.
\newblock \showarticletitle{{AIR:} Attentional Intention-Aware Recommender
  Systems}. In \bibinfo{booktitle}{\emph{35th {IEEE} International Conference
  on Data Engineering, {ICDE} 2019, Macao, China, April 8-11, 2019}}.
  \bibinfo{publisher}{{IEEE}}, \bibinfo{pages}{304--315}.
\newblock
\urldef\tempurl%
\url{https://doi.org/10.1109/ICDE.2019.00035}
\showDOI{\tempurl}


\bibitem[\protect\citeauthoryear{Chen, Yin, Nguyen, Peng, Li, and Zhou}{Chen
  et~al\mbox{.}}{2020}]%
        {ChenYNP0020}
\bibfield{author}{\bibinfo{person}{Tong Chen}, \bibinfo{person}{Hongzhi Yin},
  \bibinfo{person}{Quoc Viet~Hung Nguyen}, \bibinfo{person}{Wen{-}Chih Peng},
  \bibinfo{person}{Xue Li}, {and} \bibinfo{person}{Xiaofang Zhou}.}
  \bibinfo{year}{2020}\natexlab{}.
\newblock \showarticletitle{Sequence-Aware Factorization Machines for Temporal
  Predictive Analytics}. In \bibinfo{booktitle}{\emph{36th {IEEE} International
  Conference on Data Engineering, {ICDE} 2020, Dallas, TX, USA, April 20-24,
  2020}}. \bibinfo{publisher}{{IEEE}}, \bibinfo{pages}{1405--1416}.
\newblock


\bibitem[\protect\citeauthoryear{Chen, Yin, Zheng, Huang, Wang, and Wang}{Chen
  et~al\mbox{.}}{2021}]%
        {ChenYZHWW21}
\bibfield{author}{\bibinfo{person}{Tong Chen}, \bibinfo{person}{Hongzhi Yin},
  \bibinfo{person}{Yujia Zheng}, \bibinfo{person}{Zi Huang},
  \bibinfo{person}{Yang Wang}, {and} \bibinfo{person}{Meng Wang}.}
  \bibinfo{year}{2021}\natexlab{}.
\newblock \showarticletitle{Learning Elastic Embeddings for Customizing
  On-Device Recommenders}. In \bibinfo{booktitle}{\emph{{KDD} '21: The 27th
  {ACM} {SIGKDD} Conference on Knowledge Discovery and Data Mining, Virtual
  Event, Singapore, August 14-18, 2021}}. \bibinfo{publisher}{{ACM}},
  \bibinfo{pages}{138--147}.
\newblock


\bibitem[\protect\citeauthoryear{Chen, Xu, Zhang, Tang, Cao, Qin, and Zha}{Chen
  et~al\mbox{.}}{2018b}]%
        {chen2018sequential}
\bibfield{author}{\bibinfo{person}{Xu Chen}, \bibinfo{person}{Hongteng Xu},
  \bibinfo{person}{Yongfeng Zhang}, \bibinfo{person}{Jiaxi Tang},
  \bibinfo{person}{Yixin Cao}, \bibinfo{person}{Zheng Qin}, {and}
  \bibinfo{person}{Hongyuan Zha}.} \bibinfo{year}{2018}\natexlab{b}.
\newblock \showarticletitle{Sequential Recommendation with User Memory
  Networks}. In \bibinfo{booktitle}{\emph{Proceedings of the Eleventh {ACM}
  International Conference on Web Search and Data Mining, {WSDM} 2018, Marina
  Del Rey, CA, USA, February 5-9, 2018}}. \bibinfo{publisher}{{ACM}},
  \bibinfo{pages}{108--116}.
\newblock
\urldef\tempurl%
\url{https://doi.org/10.1145/3159652.3159668}
\showDOI{\tempurl}


\bibitem[\protect\citeauthoryear{Cheng, Yang, Lyu, and King}{Cheng
  et~al\mbox{.}}{2013}]%
        {ChengYLK13}
\bibfield{author}{\bibinfo{person}{Chen Cheng}, \bibinfo{person}{Haiqin Yang},
  \bibinfo{person}{Michael~R. Lyu}, {and} \bibinfo{person}{Irwin King}.}
  \bibinfo{year}{2013}\natexlab{}.
\newblock \showarticletitle{Where You Like to Go Next: Successive
  Point-of-Interest Recommendation}. In \bibinfo{booktitle}{\emph{{IJCAI} 2013,
  Proceedings of the 23rd International Joint Conference on Artificial
  Intelligence, Beijing, China, August 3-9, 2013}}.
  \bibinfo{publisher}{{IJCAI/AAAI}}, \bibinfo{pages}{2605--2611}.
\newblock


\bibitem[\protect\citeauthoryear{Devlin, Chang, Lee, and Toutanova}{Devlin
  et~al\mbox{.}}{2019}]%
        {DevlinCLT19}
\bibfield{author}{\bibinfo{person}{Jacob Devlin}, \bibinfo{person}{Ming{-}Wei
  Chang}, \bibinfo{person}{Kenton Lee}, {and} \bibinfo{person}{Kristina
  Toutanova}.} \bibinfo{year}{2019}\natexlab{}.
\newblock \showarticletitle{{BERT:} Pre-training of Deep Bidirectional
  Transformers for Language Understanding}. In
  \bibinfo{booktitle}{\emph{Proceedings of the 2019 Conference of the North
  American Chapter of the Association for Computational Linguistics: Human
  Language Technologies, {NAACL-HLT} 2019, Minneapolis, MN, USA, June 2-7,
  2019, Volume 1 (Long and Short Papers)}}. \bibinfo{publisher}{Association for
  Computational Linguistics}, \bibinfo{pages}{4171--4186}.
\newblock


\bibitem[\protect\citeauthoryear{Elfwing, Uchibe, and Doya}{Elfwing
  et~al\mbox{.}}{2018}]%
        {ElfwingUD18}
\bibfield{author}{\bibinfo{person}{Stefan Elfwing}, \bibinfo{person}{Eiji
  Uchibe}, {and} \bibinfo{person}{Kenji Doya}.}
  \bibinfo{year}{2018}\natexlab{}.
\newblock \showarticletitle{Sigmoid-weighted linear units for neural network
  function approximation in reinforcement learning}.
\newblock \bibinfo{journal}{\emph{Neural Networks}}  \bibinfo{volume}{107}
  (\bibinfo{year}{2018}), \bibinfo{pages}{3--11}.
\newblock


\bibitem[\protect\citeauthoryear{He, Kang, and McAuley}{He
  et~al\mbox{.}}{2017}]%
        {HeKM17}
\bibfield{author}{\bibinfo{person}{Ruining He}, \bibinfo{person}{Wang{-}Cheng
  Kang}, {and} \bibinfo{person}{Julian~J. McAuley}.}
  \bibinfo{year}{2017}\natexlab{}.
\newblock \showarticletitle{Translation-based Recommendation}. In
  \bibinfo{booktitle}{\emph{Proceedings of the Eleventh {ACM} Conference on
  Recommender Systems, RecSys 2017, Como, Italy, August 27-31, 2017}},
  \bibfield{editor}{\bibinfo{person}{Paolo Cremonesi},
  \bibinfo{person}{Francesco Ricci}, \bibinfo{person}{Shlomo Berkovsky}, {and}
  \bibinfo{person}{Alexander Tuzhilin}} (Eds.). \bibinfo{publisher}{{ACM}},
  \bibinfo{pages}{161--169}.
\newblock


\bibitem[\protect\citeauthoryear{He and McAuley}{He and McAuley}{2016}]%
        {HeM16}
\bibfield{author}{\bibinfo{person}{Ruining He} {and} \bibinfo{person}{Julian~J.
  McAuley}.} \bibinfo{year}{2016}\natexlab{}.
\newblock \showarticletitle{Fusing Similarity Models with Markov Chains for
  Sparse Sequential Recommendation}. In \bibinfo{booktitle}{\emph{{IEEE} 16th
  International Conference on Data Mining, {ICDM} 2016, December 12-15, 2016,
  Barcelona, Spain}}. \bibinfo{publisher}{{IEEE} Computer Society},
  \bibinfo{pages}{191--200}.
\newblock


\bibitem[\protect\citeauthoryear{Hendrycks and Gimpel}{Hendrycks and
  Gimpel}{2016}]%
        {HendrycksG16}
\bibfield{author}{\bibinfo{person}{Dan Hendrycks} {and} \bibinfo{person}{Kevin
  Gimpel}.} \bibinfo{year}{2016}\natexlab{}.
\newblock \showarticletitle{Bridging Nonlinearities and Stochastic Regularizers
  with Gaussian Error Linear Units}.
\newblock \bibinfo{journal}{\emph{CoRR}}  \bibinfo{volume}{abs/1606.08415}
  (\bibinfo{year}{2016}).
\newblock


\bibitem[\protect\citeauthoryear{Hidasi, Karatzoglou, Baltrunas, and
  Tikk}{Hidasi et~al\mbox{.}}{2016}]%
        {HidasiKBT15}
\bibfield{author}{\bibinfo{person}{Bal{\'{a}}zs Hidasi},
  \bibinfo{person}{Alexandros Karatzoglou}, \bibinfo{person}{Linas Baltrunas},
  {and} \bibinfo{person}{Domonkos Tikk}.} \bibinfo{year}{2016}\natexlab{}.
\newblock \showarticletitle{Session-based Recommendations with Recurrent Neural
  Networks}. In \bibinfo{booktitle}{\emph{4th International Conference on
  Learning Representations, {ICLR} 2016, San Juan, Puerto Rico, May 2-4, 2016,
  Conference Track Proceedings}}.
\newblock


\bibitem[\protect\citeauthoryear{Huang, Ren, Zhao, He, Wen, and Dong}{Huang
  et~al\mbox{.}}{2019}]%
        {HuangRZHWD19}
\bibfield{author}{\bibinfo{person}{Jin Huang}, \bibinfo{person}{Zhaochun Ren},
  \bibinfo{person}{Wayne~Xin Zhao}, \bibinfo{person}{Gaole He},
  \bibinfo{person}{Ji{-}Rong Wen}, {and} \bibinfo{person}{Daxiang Dong}.}
  \bibinfo{year}{2019}\natexlab{}.
\newblock \showarticletitle{Taxonomy-Aware Multi-Hop Reasoning Networks for
  Sequential Recommendation}. In \bibinfo{booktitle}{\emph{Proceedings of the
  Twelfth {ACM} International Conference on Web Search and Data Mining, {WSDM}
  2019, Melbourne, VIC, Australia, February 11-15, 2019}}.
  \bibinfo{publisher}{{ACM}}, \bibinfo{pages}{573--581}.
\newblock


\bibitem[\protect\citeauthoryear{Huang, Zhao, Dou, Wen, and Chang}{Huang
  et~al\mbox{.}}{2018}]%
        {HuangZDWC18}
\bibfield{author}{\bibinfo{person}{Jin Huang}, \bibinfo{person}{Wayne~Xin
  Zhao}, \bibinfo{person}{Hongjian Dou}, \bibinfo{person}{Ji{-}Rong Wen}, {and}
  \bibinfo{person}{Edward~Y. Chang}.} \bibinfo{year}{2018}\natexlab{}.
\newblock \showarticletitle{Improving Sequential Recommendation with
  Knowledge-Enhanced Memory Networks}. In \bibinfo{booktitle}{\emph{The 41st
  International {ACM} {SIGIR} Conference on Research {\&} Development in
  Information Retrieval, {SIGIR} 2018, Ann Arbor, MI, USA, July 08-12, 2018}}.
  \bibinfo{publisher}{{ACM}}, \bibinfo{pages}{505--514}.
\newblock


\bibitem[\protect\citeauthoryear{J{\'{e}}gou, Douze, and Schmid}{J{\'{e}}gou
  et~al\mbox{.}}{2011}]%
        {JegouDS11}
\bibfield{author}{\bibinfo{person}{Herv{\'{e}} J{\'{e}}gou},
  \bibinfo{person}{Matthijs Douze}, {and} \bibinfo{person}{Cordelia Schmid}.}
  \bibinfo{year}{2011}\natexlab{}.
\newblock \showarticletitle{Product Quantization for Nearest Neighbor Search}.
\newblock \bibinfo{journal}{\emph{{IEEE} Trans. Pattern Anal. Mach. Intell.}}
  (\bibinfo{year}{2011}).
\newblock


\bibitem[\protect\citeauthoryear{Kang and McAuley}{Kang and McAuley}{2018}]%
        {KangM18}
\bibfield{author}{\bibinfo{person}{Wang{-}Cheng Kang} {and}
  \bibinfo{person}{Julian~J. McAuley}.} \bibinfo{year}{2018}\natexlab{}.
\newblock \showarticletitle{Self-Attentive Sequential Recommendation}. In
  \bibinfo{booktitle}{\emph{{IEEE} International Conference on Data Mining,
  {ICDM} 2018, Singapore, November 17-20, 2018}}. \bibinfo{publisher}{{IEEE}
  Computer Society}, \bibinfo{pages}{197--206}.
\newblock


\bibitem[\protect\citeauthoryear{Kingma and Ba}{Kingma and Ba}{2015}]%
        {KingmaB14}
\bibfield{author}{\bibinfo{person}{Diederik~P. Kingma} {and}
  \bibinfo{person}{Jimmy Ba}.} \bibinfo{year}{2015}\natexlab{}.
\newblock \showarticletitle{Adam: {A} Method for Stochastic Optimization}. In
  \bibinfo{booktitle}{\emph{ICLR}}.
\newblock


\bibitem[\protect\citeauthoryear{Krichene and Rendle}{Krichene and
  Rendle}{2020}]%
        {KricheneR20}
\bibfield{author}{\bibinfo{person}{Walid Krichene} {and}
  \bibinfo{person}{Steffen Rendle}.} \bibinfo{year}{2020}\natexlab{}.
\newblock \showarticletitle{On Sampled Metrics for Item Recommendation}. In
  \bibinfo{booktitle}{\emph{KDD}}.
\newblock


\bibitem[\protect\citeauthoryear{Kumar, Zhang, and Leskovec}{Kumar
  et~al\mbox{.}}{2019}]%
        {KumarZL19}
\bibfield{author}{\bibinfo{person}{Srijan Kumar}, \bibinfo{person}{Xikun
  Zhang}, {and} \bibinfo{person}{Jure Leskovec}.}
  \bibinfo{year}{2019}\natexlab{}.
\newblock \showarticletitle{Predicting Dynamic Embedding Trajectory in Temporal
  Interaction Networks}. In \bibinfo{booktitle}{\emph{Proceedings of the 25th
  {ACM} {SIGKDD} International Conference on Knowledge Discovery {\&} Data
  Mining, {KDD} 2019, Anchorage, AK, USA, August 4-8, 2019}}.
  \bibinfo{publisher}{{ACM}}, \bibinfo{pages}{1269--1278}.
\newblock


\bibitem[\protect\citeauthoryear{Li, Chen, Luo, Yin, and Huang}{Li
  et~al\mbox{.}}{2021}]%
        {LiCLYH21}
\bibfield{author}{\bibinfo{person}{Yang Li}, \bibinfo{person}{Tong Chen},
  \bibinfo{person}{Yadan Luo}, \bibinfo{person}{Hongzhi Yin}, {and}
  \bibinfo{person}{Zi Huang}.} \bibinfo{year}{2021}\natexlab{}.
\newblock \showarticletitle{Discovering Collaborative Signals for Next {POI}
  Recommendation with Iterative Seq2Graph Augmentation}. In
  \bibinfo{booktitle}{\emph{Proceedings of the Thirtieth International Joint
  Conference on Artificial Intelligence, {IJCAI} 2021, Virtual Event /
  Montreal, Canada, 19-27 August 2021}},
  \bibfield{editor}{\bibinfo{person}{Zhi{-}Hua Zhou}} (Ed.).
  \bibinfo{publisher}{ijcai.org}, \bibinfo{pages}{1491--1497}.
\newblock


\bibitem[\protect\citeauthoryear{Li, Luo, Zhang, Sadiq, and Cui}{Li
  et~al\mbox{.}}{2019}]%
        {LiLZSC19}
\bibfield{author}{\bibinfo{person}{Yang Li}, \bibinfo{person}{Yadan Luo},
  \bibinfo{person}{Zheng Zhang}, \bibinfo{person}{Shazia~W. Sadiq}, {and}
  \bibinfo{person}{Peng Cui}.} \bibinfo{year}{2019}\natexlab{}.
\newblock \showarticletitle{Context-Aware Attention-Based Data Augmentation for
  {POI} Recommendation}. In \bibinfo{booktitle}{\emph{35th {IEEE} International
  Conference on Data Engineering Workshops, {ICDE} Workshops 2019, Macao,
  China, April 8-12, 2019}}. \bibinfo{publisher}{{IEEE}},
  \bibinfo{pages}{177--184}.
\newblock


\bibitem[\protect\citeauthoryear{Lian, Wang, Liu, Lian, Chen, and Xie}{Lian
  et~al\mbox{.}}{2020}]%
        {LianWLLC020}
\bibfield{author}{\bibinfo{person}{Defu Lian}, \bibinfo{person}{Haoyu Wang},
  \bibinfo{person}{Zheng Liu}, \bibinfo{person}{Jianxun Lian},
  \bibinfo{person}{Enhong Chen}, {and} \bibinfo{person}{Xing Xie}.}
  \bibinfo{year}{2020}\natexlab{}.
\newblock \showarticletitle{LightRec: {A} Memory and Search-Efficient
  Recommender System}. In \bibinfo{booktitle}{\emph{{WWW} '20: The Web
  Conference 2020, Taipei, Taiwan, April 20-24, 2020}},
  \bibfield{editor}{\bibinfo{person}{Yennun Huang}, \bibinfo{person}{Irwin
  King}, \bibinfo{person}{Tie{-}Yan Liu}, {and} \bibinfo{person}{Maarten van
  Steen}} (Eds.). \bibinfo{publisher}{{ACM} / {IW3C2}},
  \bibinfo{pages}{695--705}.
\newblock


\bibitem[\protect\citeauthoryear{Liu, Zhao, Wang, Liu, and Tang}{Liu
  et~al\mbox{.}}{2020}]%
        {liu2020automated}
\bibfield{author}{\bibinfo{person}{Haochen Liu}, \bibinfo{person}{Xiangyu
  Zhao}, \bibinfo{person}{Chong Wang}, \bibinfo{person}{Xiaobing Liu}, {and}
  \bibinfo{person}{Jiliang Tang}.} \bibinfo{year}{2020}\natexlab{}.
\newblock \showarticletitle{Automated Embedding Size Search in Deep Recommender
  Systems}. In \bibinfo{booktitle}{\emph{SIGIR}}. \bibinfo{pages}{2307--2316}.
\newblock


\bibitem[\protect\citeauthoryear{Liu, Gao, Chen, Jin, and Li}{Liu
  et~al\mbox{.}}{2021}]%
        {LiuGCJL21}
\bibfield{author}{\bibinfo{person}{Siyi Liu}, \bibinfo{person}{Chen Gao},
  \bibinfo{person}{Yihong Chen}, \bibinfo{person}{Depeng Jin}, {and}
  \bibinfo{person}{Yong Li}.} \bibinfo{year}{2021}\natexlab{}.
\newblock \showarticletitle{Learnable Embedding sizes for Recommender Systems}.
  In \bibinfo{booktitle}{\emph{9th International Conference on Learning
  Representations, {ICLR} 2021, Virtual Event, Austria, May 3-7, 2021}}.
  \bibinfo{publisher}{OpenReview.net}.
\newblock


\bibitem[\protect\citeauthoryear{Luo, Huang, Li, Shen, Yang, and Cui}{Luo
  et~al\mbox{.}}{2020}]%
        {YadanLowBit20}
\bibfield{author}{\bibinfo{person}{Yadan Luo}, \bibinfo{person}{Zi Huang},
  \bibinfo{person}{Yang Li}, \bibinfo{person}{Fumin Shen},
  \bibinfo{person}{Yang Yang}, {and} \bibinfo{person}{Peng Cui}.}
  \bibinfo{year}{2020}\natexlab{}.
\newblock \showarticletitle{Collaborative Learning for Extremely Low Bit
  Asymmetric Hashing}.
\newblock \bibinfo{journal}{\emph{IEEE Transactions on Knowledge and Data
  Engineering}} (\bibinfo{year}{2020}), \bibinfo{pages}{1--1}.
\newblock


\bibitem[\protect\citeauthoryear{Lv, Jin, Yu, Sun, Lin, Yang, and Ng}{Lv
  et~al\mbox{.}}{2019}]%
        {LvJYSLYN19}
\bibfield{author}{\bibinfo{person}{Fuyu Lv}, \bibinfo{person}{Taiwei Jin},
  \bibinfo{person}{Changlong Yu}, \bibinfo{person}{Fei Sun},
  \bibinfo{person}{Quan Lin}, \bibinfo{person}{Keping Yang}, {and}
  \bibinfo{person}{Wilfred Ng}.} \bibinfo{year}{2019}\natexlab{}.
\newblock \showarticletitle{{SDM:} Sequential Deep Matching Model for Online
  Large-scale Recommender System}. In \bibinfo{booktitle}{\emph{Proceedings of
  the 28th {ACM} International Conference on Information and Knowledge
  Management, {CIKM} 2019, Beijing, China, November 3-7, 2019}}.
  \bibinfo{publisher}{{ACM}}, \bibinfo{pages}{2635--2643}.
\newblock


\bibitem[\protect\citeauthoryear{Qiu, Huang, Chen, and Yin}{Qiu
  et~al\mbox{.}}{2021}]%
        {QiuPosRec21}
\bibfield{author}{\bibinfo{person}{Ruihong Qiu}, \bibinfo{person}{Zi Huang},
  \bibinfo{person}{Tong Chen}, {and} \bibinfo{person}{Hongzhi Yin}.}
  \bibinfo{year}{2021}\natexlab{}.
\newblock \showarticletitle{Exploiting Positional Information for Session-based
  Recommendation}.
\newblock \bibinfo{journal}{\emph{CoRR}}  \bibinfo{volume}{abs/2107.00846}
  (\bibinfo{year}{2021}).
\newblock


\bibitem[\protect\citeauthoryear{Qiu, Huang, Li, and Yin}{Qiu
  et~al\mbox{.}}{2020a}]%
        {QiuHLY20}
\bibfield{author}{\bibinfo{person}{Ruihong Qiu}, \bibinfo{person}{Zi Huang},
  \bibinfo{person}{Jingjing Li}, {and} \bibinfo{person}{Hongzhi Yin}.}
  \bibinfo{year}{2020}\natexlab{a}.
\newblock \showarticletitle{Exploiting Cross-session Information for
  Session-based Recommendation with Graph Neural Networks}.
\newblock \bibinfo{journal}{\emph{{ACM} Trans. Inf. Syst.}}
  \bibinfo{volume}{38}, \bibinfo{number}{3} (\bibinfo{year}{2020}),
  \bibinfo{pages}{22:1--22:23}.
\newblock


\bibitem[\protect\citeauthoryear{Qiu, Li, Huang, and Yin}{Qiu
  et~al\mbox{.}}{2019}]%
        {QiuLHY19}
\bibfield{author}{\bibinfo{person}{Ruihong Qiu}, \bibinfo{person}{Jingjing Li},
  \bibinfo{person}{Zi Huang}, {and} \bibinfo{person}{Hongzhi Yin}.}
  \bibinfo{year}{2019}\natexlab{}.
\newblock \showarticletitle{Rethinking the Item Order in Session-based
  Recommendation with Graph Neural Networks}. In
  \bibinfo{booktitle}{\emph{Proceedings of the 28th {ACM} International
  Conference on Information and Knowledge Management, {CIKM} 2019, Beijing,
  China, November 3-7, 2019}}. \bibinfo{publisher}{{ACM}},
  \bibinfo{pages}{579--588}.
\newblock


\bibitem[\protect\citeauthoryear{Qiu, Yin, Huang, and Chen}{Qiu
  et~al\mbox{.}}{2020b}]%
        {QiuYHC20}
\bibfield{author}{\bibinfo{person}{Ruihong Qiu}, \bibinfo{person}{Hongzhi Yin},
  \bibinfo{person}{Zi Huang}, {and} \bibinfo{person}{Tong Chen}.}
  \bibinfo{year}{2020}\natexlab{b}.
\newblock \showarticletitle{{GAG:} Global Attributed Graph Neural Network for
  Streaming Session-based Recommendation}. In
  \bibinfo{booktitle}{\emph{Proceedings of the 43rd International {ACM} {SIGIR}
  conference on research and development in Information Retrieval, {SIGIR}
  2020, Virtual Event, China, July 25-30, 2020}}. \bibinfo{publisher}{{ACM}},
  \bibinfo{pages}{669--678}.
\newblock


\bibitem[\protect\citeauthoryear{Ren, Chen, Li, Ren, Ma, and de~Rijke}{Ren
  et~al\mbox{.}}{2019}]%
        {RenCLR0R19}
\bibfield{author}{\bibinfo{person}{Pengjie Ren}, \bibinfo{person}{Zhumin Chen},
  \bibinfo{person}{Jing Li}, \bibinfo{person}{Zhaochun Ren},
  \bibinfo{person}{Jun Ma}, {and} \bibinfo{person}{Maarten de Rijke}.}
  \bibinfo{year}{2019}\natexlab{}.
\newblock \showarticletitle{RepeatNet: {A} Repeat Aware Neural Recommendation
  Machine for Session-Based Recommendation}. In \bibinfo{booktitle}{\emph{The
  Thirty-Third {AAAI} Conference on Artificial Intelligence, {AAAI} 2019, The
  Thirty-First Innovative Applications of Artificial Intelligence Conference,
  {IAAI} 2019, The Ninth {AAAI} Symposium on Educational Advances in Artificial
  Intelligence, {EAAI} 2019, Honolulu, Hawaii, USA, January 27 - February 1,
  2019}}. \bibinfo{publisher}{{AAAI} Press}, \bibinfo{pages}{4806--4813}.
\newblock


\bibitem[\protect\citeauthoryear{Rendle, Freudenthaler, and
  Schmidt{-}Thieme}{Rendle et~al\mbox{.}}{2010}]%
        {RendleFS10}
\bibfield{author}{\bibinfo{person}{Steffen Rendle}, \bibinfo{person}{Christoph
  Freudenthaler}, {and} \bibinfo{person}{Lars Schmidt{-}Thieme}.}
  \bibinfo{year}{2010}\natexlab{}.
\newblock \showarticletitle{Factorizing personalized Markov chains for
  next-basket recommendation}. In \bibinfo{booktitle}{\emph{Proceedings of the
  19th International Conference on World Wide Web, {WWW} 2010, Raleigh, North
  Carolina, USA, April 26-30, 2010}}. \bibinfo{publisher}{{ACM}},
  \bibinfo{pages}{811--820}.
\newblock


\bibitem[\protect\citeauthoryear{Shi, Mudigere, Naumov, and Yang}{Shi
  et~al\mbox{.}}{2020}]%
        {ShiMNY20}
\bibfield{author}{\bibinfo{person}{Hao{-}Jun~Michael Shi},
  \bibinfo{person}{Dheevatsa Mudigere}, \bibinfo{person}{Maxim Naumov}, {and}
  \bibinfo{person}{Jiyan Yang}.} \bibinfo{year}{2020}\natexlab{}.
\newblock \showarticletitle{Compositional Embeddings Using Complementary
  Partitions for Memory-Efficient Recommendation Systems}. In
  \bibinfo{booktitle}{\emph{{KDD} '20: The 26th {ACM} {SIGKDD} Conference on
  Knowledge Discovery and Data Mining, Virtual Event, CA, USA, August 23-27,
  2020}}. \bibinfo{publisher}{{ACM}}, \bibinfo{pages}{165--175}.
\newblock


\bibitem[\protect\citeauthoryear{Shi, Cao, Zhang, Li, and Xu}{Shi
  et~al\mbox{.}}{2016}]%
        {shi2016edge}
\bibfield{author}{\bibinfo{person}{Weisong Shi}, \bibinfo{person}{Jie Cao},
  \bibinfo{person}{Quan Zhang}, \bibinfo{person}{Youhuizi Li}, {and}
  \bibinfo{person}{Lanyu Xu}.} \bibinfo{year}{2016}\natexlab{}.
\newblock \showarticletitle{Edge computing: Vision and challenges}.
\newblock \bibinfo{journal}{\emph{IEEE internet of things journal}}
  \bibinfo{volume}{3}, \bibinfo{number}{5} (\bibinfo{year}{2016}),
  \bibinfo{pages}{637--646}.
\newblock


\bibitem[\protect\citeauthoryear{Shu and Nakayama}{Shu and Nakayama}{2018}]%
        {ShuN18}
\bibfield{author}{\bibinfo{person}{Raphael Shu} {and} \bibinfo{person}{Hideki
  Nakayama}.} \bibinfo{year}{2018}\natexlab{}.
\newblock \showarticletitle{Compressing Word Embeddings via Deep Compositional
  Code Learning}. In \bibinfo{booktitle}{\emph{6th International Conference on
  Learning Representations, {ICLR} 2018, Vancouver, BC, Canada, April 30 - May
  3, 2018, Conference Track Proceedings}}. \bibinfo{publisher}{OpenReview.net}.
\newblock


\bibitem[\protect\citeauthoryear{Sun, Liu, Wu, Pei, Lin, Ou, and Jiang}{Sun
  et~al\mbox{.}}{2019}]%
        {SunLWPLOJ19}
\bibfield{author}{\bibinfo{person}{Fei Sun}, \bibinfo{person}{Jun Liu},
  \bibinfo{person}{Jian Wu}, \bibinfo{person}{Changhua Pei},
  \bibinfo{person}{Xiao Lin}, \bibinfo{person}{Wenwu Ou}, {and}
  \bibinfo{person}{Peng Jiang}.} \bibinfo{year}{2019}\natexlab{}.
\newblock \showarticletitle{BERT4Rec: Sequential Recommendation with
  Bidirectional Encoder Representations from Transformer}. In
  \bibinfo{booktitle}{\emph{Proceedings of the 28th {ACM} International
  Conference on Information and Knowledge Management, {CIKM} 2019, Beijing,
  China, November 3-7, 2019}}. \bibinfo{publisher}{{ACM}},
  \bibinfo{pages}{1441--1450}.
\newblock


\bibitem[\protect\citeauthoryear{Sun, Qian, Chen, Liang, Nguyen, and Yin}{Sun
  et~al\mbox{.}}{2020}]%
        {SunQCLNY20}
\bibfield{author}{\bibinfo{person}{Ke Sun}, \bibinfo{person}{Tieyun Qian},
  \bibinfo{person}{Tong Chen}, \bibinfo{person}{Yile Liang},
  \bibinfo{person}{Quoc Viet~Hung Nguyen}, {and} \bibinfo{person}{Hongzhi
  Yin}.} \bibinfo{year}{2020}\natexlab{}.
\newblock \showarticletitle{Where to Go Next: Modeling Long- and Short-Term
  User Preferences for Point-of-Interest Recommendation}. In
  \bibinfo{booktitle}{\emph{The Thirty-Fourth {AAAI} Conference on Artificial
  Intelligence, {AAAI} 2020, The Thirty-Second Innovative Applications of
  Artificial Intelligence Conference, {IAAI} 2020, The Tenth {AAAI} Symposium
  on Educational Advances in Artificial Intelligence, {EAAI} 2020, New York,
  NY, USA, February 7-12, 2020}}. \bibinfo{publisher}{{AAAI} Press},
  \bibinfo{pages}{214--221}.
\newblock


\bibitem[\protect\citeauthoryear{Tang and Wang}{Tang and Wang}{2018}]%
        {TangW18}
\bibfield{author}{\bibinfo{person}{Jiaxi Tang} {and} \bibinfo{person}{Ke
  Wang}.} \bibinfo{year}{2018}\natexlab{}.
\newblock \showarticletitle{Personalized Top-N Sequential Recommendation via
  Convolutional Sequence Embedding}. In \bibinfo{booktitle}{\emph{Proceedings
  of the Eleventh {ACM} International Conference on Web Search and Data Mining,
  {WSDM} 2018, Marina Del Rey, CA, USA, February 5-9, 2018}}.
  \bibinfo{publisher}{{ACM}}, \bibinfo{pages}{565--573}.
\newblock


\bibitem[\protect\citeauthoryear{Vaswani, Shazeer, Parmar, Uszkoreit, Jones,
  Gomez, Kaiser, and Polosukhin}{Vaswani et~al\mbox{.}}{2017}]%
        {VaswaniSPUJGKP17}
\bibfield{author}{\bibinfo{person}{Ashish Vaswani}, \bibinfo{person}{Noam
  Shazeer}, \bibinfo{person}{Niki Parmar}, \bibinfo{person}{Jakob Uszkoreit},
  \bibinfo{person}{Llion Jones}, \bibinfo{person}{Aidan~N. Gomez},
  \bibinfo{person}{Lukasz Kaiser}, {and} \bibinfo{person}{Illia Polosukhin}.}
  \bibinfo{year}{2017}\natexlab{}.
\newblock \showarticletitle{Attention is All you Need}. In
  \bibinfo{booktitle}{\emph{Advances in Neural Information Processing Systems
  30: Annual Conference on Neural Information Processing Systems 2017, December
  4-9, 2017, Long Beach, CA, {USA}}}. \bibinfo{pages}{5998--6008}.
\newblock


\bibitem[\protect\citeauthoryear{Wang, Huang, Zhao, Zhang, Zhao, and Lee}{Wang
  et~al\mbox{.}}{2018}]%
        {wang2018billion}
\bibfield{author}{\bibinfo{person}{Jizhe Wang}, \bibinfo{person}{Pipei Huang},
  \bibinfo{person}{Huan Zhao}, \bibinfo{person}{Zhibo Zhang},
  \bibinfo{person}{Binqiang Zhao}, {and} \bibinfo{person}{Dik~Lun Lee}.}
  \bibinfo{year}{2018}\natexlab{}.
\newblock \showarticletitle{Billion-scale commodity embedding for e-commerce
  recommendation in alibaba}. In \bibinfo{booktitle}{\emph{SIGKDD}}.
  \bibinfo{pages}{839--848}.
\newblock


\bibitem[\protect\citeauthoryear{Wang, Yin, Chen, Huang, Wang, Zhao, and
  Hung}{Wang et~al\mbox{.}}{2020}]%
        {WangYCHWZH20}
\bibfield{author}{\bibinfo{person}{Qinyong Wang}, \bibinfo{person}{Hongzhi
  Yin}, \bibinfo{person}{Tong Chen}, \bibinfo{person}{Zi Huang},
  \bibinfo{person}{Hao Wang}, \bibinfo{person}{Yanchang Zhao}, {and}
  \bibinfo{person}{Nguyen Quoc~Viet Hung}.} \bibinfo{year}{2020}\natexlab{}.
\newblock \showarticletitle{Next Point-of-Interest Recommendation on
  Resource-Constrained Mobile Devices}. In \bibinfo{booktitle}{\emph{{WWW} '20:
  The Web Conference 2020, Taipei, Taiwan, April 20-24, 2020}}.
  \bibinfo{publisher}{{ACM} / {IW3C2}}, \bibinfo{pages}{906--916}.
\newblock


\bibitem[\protect\citeauthoryear{Wu, Ahmed, Beutel, Smola, and Jing}{Wu
  et~al\mbox{.}}{2017}]%
        {WuABSJ17}
\bibfield{author}{\bibinfo{person}{Chao{-}Yuan Wu}, \bibinfo{person}{Amr
  Ahmed}, \bibinfo{person}{Alex Beutel}, \bibinfo{person}{Alexander~J. Smola},
  {and} \bibinfo{person}{How Jing}.} \bibinfo{year}{2017}\natexlab{}.
\newblock \showarticletitle{Recurrent Recommender Networks}. In
  \bibinfo{booktitle}{\emph{Proceedings of the Tenth {ACM} International
  Conference on Web Search and Data Mining, {WSDM} 2017, Cambridge, United
  Kingdom, February 6-10, 2017}}. \bibinfo{publisher}{{ACM}},
  \bibinfo{pages}{495--503}.
\newblock


\bibitem[\protect\citeauthoryear{Wu, Fan, Baevski, Dauphin, and Auli}{Wu
  et~al\mbox{.}}{2019a}]%
        {wu2019pay}
\bibfield{author}{\bibinfo{person}{Felix Wu}, \bibinfo{person}{Angela Fan},
  \bibinfo{person}{Alexei Baevski}, \bibinfo{person}{Yann Dauphin}, {and}
  \bibinfo{person}{Michael Auli}.} \bibinfo{year}{2019}\natexlab{a}.
\newblock \showarticletitle{Pay Less Attention with Lightweight and Dynamic
  Convolutions}. In \bibinfo{booktitle}{\emph{ICLR}}.
\newblock


\bibitem[\protect\citeauthoryear{Wu, Fan, Baevski, Dauphin, and Auli}{Wu
  et~al\mbox{.}}{2019b}]%
        {WuFBDA19}
\bibfield{author}{\bibinfo{person}{Felix Wu}, \bibinfo{person}{Angela Fan},
  \bibinfo{person}{Alexei Baevski}, \bibinfo{person}{Yann~N. Dauphin}, {and}
  \bibinfo{person}{Michael Auli}.} \bibinfo{year}{2019}\natexlab{b}.
\newblock \showarticletitle{Pay Less Attention with Lightweight and Dynamic
  Convolutions}. In \bibinfo{booktitle}{\emph{7th International Conference on
  Learning Representations, {ICLR} 2019, New Orleans, LA, USA, May 6-9, 2019}}.
  \bibinfo{publisher}{OpenReview.net}.
\newblock


\bibitem[\protect\citeauthoryear{Wu, Liu, Lin, Lin, and Han}{Wu
  et~al\mbox{.}}{2020}]%
        {WuLLLH20}
\bibfield{author}{\bibinfo{person}{Zhanghao Wu}, \bibinfo{person}{Zhijian Liu},
  \bibinfo{person}{Ji Lin}, \bibinfo{person}{Yujun Lin}, {and}
  \bibinfo{person}{Song Han}.} \bibinfo{year}{2020}\natexlab{}.
\newblock \showarticletitle{Lite Transformer with Long-Short Range Attention}.
  In \bibinfo{booktitle}{\emph{8th International Conference on Learning
  Representations, {ICLR} 2020, Addis Ababa, Ethiopia, April 26-30, 2020}}.
  \bibinfo{publisher}{OpenReview.net}.
\newblock


\bibitem[\protect\citeauthoryear{Xin, Karatzoglou, Arapakis, and Jose}{Xin
  et~al\mbox{.}}{2020}]%
        {XinKAJ20}
\bibfield{author}{\bibinfo{person}{Xin Xin}, \bibinfo{person}{Alexandros
  Karatzoglou}, \bibinfo{person}{Ioannis Arapakis}, {and}
  \bibinfo{person}{Joemon~M. Jose}.} \bibinfo{year}{2020}\natexlab{}.
\newblock \showarticletitle{Self-Supervised Reinforcement Learning for
  Recommender Systems}. In \bibinfo{booktitle}{\emph{Proceedings of the 43rd
  International {ACM} {SIGIR} conference on research and development in
  Information Retrieval, {SIGIR} 2020, Virtual Event, China, July 25-30,
  2020}}. \bibinfo{publisher}{{ACM}}, \bibinfo{pages}{931--940}.
\newblock


\bibitem[\protect\citeauthoryear{Yang, Yang, Dyer, He, Smola, and Hovy}{Yang
  et~al\mbox{.}}{2016}]%
        {YangYDHSH16}
\bibfield{author}{\bibinfo{person}{Zichao Yang}, \bibinfo{person}{Diyi Yang},
  \bibinfo{person}{Chris Dyer}, \bibinfo{person}{Xiaodong He},
  \bibinfo{person}{Alexander~J. Smola}, {and} \bibinfo{person}{Eduard~H.
  Hovy}.} \bibinfo{year}{2016}\natexlab{}.
\newblock \showarticletitle{Hierarchical Attention Networks for Document
  Classification}. In \bibinfo{booktitle}{\emph{NAACL}}.
  \bibinfo{publisher}{The Association for Computational Linguistics},
  \bibinfo{pages}{1480--1489}.
\newblock


\bibitem[\protect\citeauthoryear{Yin and Cui}{Yin and Cui}{2016}]%
        {YinC16}
\bibfield{author}{\bibinfo{person}{Hongzhi Yin} {and} \bibinfo{person}{Bin
  Cui}.} \bibinfo{year}{2016}\natexlab{}.
\newblock \bibinfo{booktitle}{\emph{Spatio-Temporal Recommendation in Social
  Media}}.
\newblock \bibinfo{publisher}{Springer}.
\newblock


\bibitem[\protect\citeauthoryear{Zhang, Li, Huang, and Xu}{Zhang
  et~al\mbox{.}}{2021}]%
        {ZhangHash21}
\bibfield{author}{\bibinfo{person}{Peng-Fei Zhang}, \bibinfo{person}{Yang Li},
  \bibinfo{person}{Zi Huang}, {and} \bibinfo{person}{Xin-Shun Xu}.}
  \bibinfo{year}{2021}\natexlab{}.
\newblock \showarticletitle{Aggregation-based Graph Convolutional Hashing for
  Unsupervised Cross-modal Retrieval}.
\newblock \bibinfo{journal}{\emph{IEEE Transactions on Multimedia}}
  (\bibinfo{year}{2021}), \bibinfo{pages}{1--1}.
\newblock


\bibitem[\protect\citeauthoryear{Zhang, Yin, Huang, Du, Yang, and Lian}{Zhang
  et~al\mbox{.}}{2018}]%
        {ZhangYHDYL18}
\bibfield{author}{\bibinfo{person}{Yan Zhang}, \bibinfo{person}{Hongzhi Yin},
  \bibinfo{person}{Zi Huang}, \bibinfo{person}{Xingzhong Du},
  \bibinfo{person}{Guowu Yang}, {and} \bibinfo{person}{Defu Lian}.}
  \bibinfo{year}{2018}\natexlab{}.
\newblock \showarticletitle{Discrete Deep Learning for Fast Content-Aware
  Recommendation}. In \bibinfo{booktitle}{\emph{Proceedings of the Eleventh
  {ACM} International Conference on Web Search and Data Mining, {WSDM} 2018,
  Marina Del Rey, CA, USA, February 5-9, 2018}}. \bibinfo{publisher}{{ACM}},
  \bibinfo{pages}{717--726}.
\newblock


\bibitem[\protect\citeauthoryear{Zhang, Xie, Li, Li, and Huang}{Zhang
  et~al\mbox{.}}{2019}]%
        {ZhangXLLH19}
\bibfield{author}{\bibinfo{person}{Zheng Zhang}, \bibinfo{person}{Guo{-}Sen
  Xie}, \bibinfo{person}{Yang Li}, \bibinfo{person}{Sheng Li}, {and}
  \bibinfo{person}{Zi Huang}.} \bibinfo{year}{2019}\natexlab{}.
\newblock \showarticletitle{{SADIH:} Semantic-Aware DIscrete Hashing}. In
  \bibinfo{booktitle}{\emph{The Thirty-Third {AAAI} Conference on Artificial
  Intelligence, {AAAI} 2019, The Thirty-First Innovative Applications of
  Artificial Intelligence Conference, {IAAI} 2019, The Ninth {AAAI} Symposium
  on Educational Advances in Artificial Intelligence, {EAAI} 2019, Honolulu,
  Hawaii, USA, January 27 - February 1, 2019}}. \bibinfo{publisher}{{AAAI}
  Press}, \bibinfo{pages}{5853--5860}.
\newblock


\bibitem[\protect\citeauthoryear{Zheng, Guo, Chen, Yu, and Jiang}{Zheng
  et~al\mbox{.}}{2020}]%
        {zheng2020sentiment}
\bibfield{author}{\bibinfo{person}{Lin Zheng}, \bibinfo{person}{Naicheng Guo},
  \bibinfo{person}{Weihao Chen}, \bibinfo{person}{Jin Yu}, {and}
  \bibinfo{person}{Dazhi Jiang}.} \bibinfo{year}{2020}\natexlab{}.
\newblock \showarticletitle{Sentiment-guided Sequential Recommendation}. In
  \bibinfo{booktitle}{\emph{Proceedings of the 43rd International {ACM} {SIGIR}
  conference on research and development in Information Retrieval, {SIGIR}
  2020, Virtual Event, China, July 25-30, 2020}}. \bibinfo{publisher}{{ACM}},
  \bibinfo{pages}{1957--1960}.
\newblock


\bibitem[\protect\citeauthoryear{Zhou, Alvarez, and Porikli}{Zhou
  et~al\mbox{.}}{2016}]%
        {ZhouAP16}
\bibfield{author}{\bibinfo{person}{Hao Zhou}, \bibinfo{person}{Jose~M.
  Alvarez}, {and} \bibinfo{person}{Fatih Porikli}.}
  \bibinfo{year}{2016}\natexlab{}.
\newblock \showarticletitle{Less Is More: Towards Compact CNNs}. In
  \bibinfo{booktitle}{\emph{Computer Vision - {ECCV} 2016 - 14th European
  Conference, Amsterdam, The Netherlands, October 11-14, 2016, Proceedings,
  Part {IV}}} \emph{(\bibinfo{series}{Lecture Notes in Computer Science},
  Vol.~\bibinfo{volume}{9908})}. \bibinfo{publisher}{Springer},
  \bibinfo{pages}{662--677}.
\newblock


\bibitem[\protect\citeauthoryear{Zhou, Wang, Zhao, Zhu, Wang, Zhang, Wang, and
  Wen}{Zhou et~al\mbox{.}}{2020}]%
        {zhou2020s3}
\bibfield{author}{\bibinfo{person}{Kun Zhou}, \bibinfo{person}{Hui Wang},
  \bibinfo{person}{Wayne~Xin Zhao}, \bibinfo{person}{Yutao Zhu},
  \bibinfo{person}{Sirui Wang}, \bibinfo{person}{Fuzheng Zhang},
  \bibinfo{person}{Zhongyuan Wang}, {and} \bibinfo{person}{Ji{-}Rong Wen}.}
  \bibinfo{year}{2020}\natexlab{}.
\newblock \showarticletitle{S3-Rec: Self-Supervised Learning for Sequential
  Recommendation with Mutual Information Maximization}. In
  \bibinfo{booktitle}{\emph{{CIKM} '20: The 29th {ACM} International Conference
  on Information and Knowledge Management, Virtual Event, Ireland, October
  19-23, 2020}}. \bibinfo{publisher}{{ACM}}, \bibinfo{pages}{1893--1902}.
\newblock


\end{thebibliography}


\end{document}